\documentclass[usenatbib,useAMS,usegraphicx]{mn2e}
\usepackage{amsmath}
\usepackage{bm}
\usepackage{amsmath}
\usepackage{mathrsfs}

\newcommand\beq{\begin{equation}}
\newcommand\eeq{\end{equation}}

\newcommand\bdm{\begin{displaymath}}
\newcommand\edm{\end{displaymath}}

\begin{document}

\title[Hierarchical mass modelling]{A Robust Determination of Milky Way Satellite Properties using hierarchical mass modelling}

\author[Martinez]
{Gregory D. Martinez$^1$\thanks{E-mail:
gregory.martinez@fysik.su.se}\\
$^1$ The Oskar Klein Center, Department of Physics, Stockholm University,  Albanova, SE-10691 Stockholm, Sweden
\\
}
\maketitle

\date{June 13, 2015}
\pagerange{2524--2535}\pubyear{2015}\volume{451}
\begin{abstract}
We introduce a new methodology to robustly determine the mass profile, as well as the overall distribution, of Local Group satellite galaxies.  Specifically we employ a statistical multilevel modelling technique, Bayesian hierarchical modelling, to simultaneously constrain the properties of individual Local Group Milky Way satellite galaxies and the characteristics of the Milky Way satellite population.  We show that this methodology reduces the uncertainty in individual dwarf galaxy mass measurements up to a factor of a few for the faintest galaxies.  We find that the distribution of Milky Way satellites inferred by this analysis, with the exception of the apparent lack of high-mass haloes, is consistent with the $\Lambda$ cold dark matter ($\Lambda$CDM) paradigm.  In particular we find that both the measured relationship between the maximum circular velocity and the radius at this velocity, as well as the inferred relationship between the mass within $300$pc and luminosity, match the values predicted by $\Lambda$CDM simulations for halos with maximum circular velocities below $20$ km s$^{-1}$.  Perhaps more striking is that this analysis seems to suggest a more cusped ``average'' halo shape that is shared by these galaxies.  While this study reconciles many of the observed properties of the Milky Way satellite distribution with that of $\Lambda$CDM simulations, we find that there is still a deficit of satellites with maximum circular velocities of $20$-$40$ km s$^{-1}$.
\end{abstract}

\section{Introduction}

The $\Lambda$ cold dark matter ($\Lambda$CDM) paradigm makes far-reaching predictions about galaxy formation and cosmology.  Observations of the CMB and large-scale structure have confirmed many of these predictions \citep{Efstathiou1992, Riess1998, Komatsu2011} making this paradigm the favoured cosmological model.  Although $\Lambda$CDM has enjoyed much success at large scales, there are some indications of discrepancies at smaller scales.  These discrepancies can be categorized in two ways:  individual halo density profiles and overall distributions of sub-structure.  Particularly, it has been suggested that the Milky Way Local Group dwarf spheroidal (dSph) galaxies have a much flatter mass--luminosity relation and possess much shallower inner density profiles than those predicted by $\Lambda$CDM \citep{Goerdt2006, Gilmore2007, Evans2009, Walker2011, Agnello2012}.  This has motivated development of alternative hypotheses such as warm and self-interacting dark matter \citep{Bond1980, SpergelSteinhardt2000, Dalcanton2001, Strigari2006, Boyarsky2009, Carlson1992, Burkert2000, Colin2002, Ahn2005, LoebWeiner2011, Macci2012, Rocha2013}. Many of these models give flatter inner density profiles at dSph scales while retaining CDM's successful large-scale predictions.  However, recent studies have suggested that shallow inner density profiles may be a consequence of the flattening of inner cusps from baryonic effects \citep{Mashchenko2006, Mashchenko2008, Arraki2012, Brooks2012, Governato2012, Parry2012, Pontzen2012, Zolotov2012}.  

Perhaps just as troubling is the apparent disagreement between the sub-structure distributions observed locally and those predicted by $\Lambda$CDM simulations.  One of the best studied discrepancies is the apparent lack of numerous low-mass sub-structures predicted by the $\Lambda$CDM paradigm \citep{Kauffmann1993, Klypin1999, Moore1999, Bullock2010}.  This so-called ``Missing Satellites Problem'' persisted even after the discovery of several additional ultra-faint satellite galaxies \citep{Grebel2000, Willman2005, Irwin2007, Simon2007, Belokurov2008, Liu2008, Martin2008, Watkins2009, Belokurov2010, Martinez2011, Simon2011}.   This problem is further complicated by the fact that these galaxies may share a common mass scale over several orders of magnitude in luminosity \citep{Strigari2007, Strigari2008}.  Many of these discrepancies have been addressed by including astrophysical and observational effects, such as suppression of star formation due to reionization and feedback \citep{Quinn1996, Thoul1996, Navarro1997a, Barkana1999, Klypin1999, Bullock2000, Gnedin2000, Benson2002, Hoeft2006, Madau2008, Alvarez2009}, and selection biases \citep{Willman2004, Simon2007, Tollerud2008, Koposov2009, Walsh2009, Rashkov2012}.  However, a slight correlation between the mass and luminosity is still common in these improved dark matter analyses \citep{Ricotti2005, Koposov2009, Maccio2009, Okamoto2009, Busha2010, Font2011, Rashkov2012}.  To make matters worse, there is also a deficit of observed satellites at the high-mass end of the dwarf satellite mass spectrum.  Dubbed the ``Too Big to Fail'' problem, \citet[also see \citealt{Vera-Ciro2013}]{Boylan-Kolchin2011} noted that the Aquarius simulations predict at least $10$ sub-haloes with a maximum circular velocity greater than $25$ km/sec. Attempts to place the most luminous known satellites into haloes of this size result in halo densities inconsistent with CDM simulations \citep{Boylan-Kolchin2012}.  One possible solution to this discrepancy may originate in the same baryonic processes used to explain the cusp/core problem -- in that core-like central regions created by supernova feedback and tidal stripping make these galaxies more suspectible to disruption by the Milky Way disc \citep{Penarrubia2010, DiCintio2011, Brooks2012, Zolotov2012, Penarrubia2012}.  On the other hand, if the mass of the Milky Way halo has been overestimated, this apparent lack of high-mass sub-haloes may be due to a statistical anomaly \citep{Wang2012}.  However, the efficacy of these solutions has been disputed by various authors \citep{Boylan-Kolchin2012, Strigari2012, Garrison-Kimmel2013}.  For a general review of these issues see \citet{Strigari2012-review}.

While much effort has been focused on reconciling CDM predictions with various dwarf galaxy observations, little attention has been paid to the statistical consistency of the measurements themselves.  Current Jeans modelling methods used to constrain dSphs halo properties are dominated by assumed prior probabilities \citep{Martinez2009, Walker2009a, Wolf2010, Walker2012}.  This can potentially be overcome by using more advanced methods such as phase space, Schwarzschild, or higher-order Jeans modelling \citep{Lokas2005, Wu2007, Amorisco2012, Jardel2012, Richardson2012, Breddels2013, Jardel2013}.  However, these methodologies introduce systematics such as binary contributions and membership effects that, to date, have not been included in such analyses \citep{Walker2009, Minor2010, Martinez2011}.  Therefore, prior dominance is crucial to this discussion as it not only affects the characterization of individual dSphs, but has an unknown effect on the inferred parameters of the population.  In this paper we aim to address this issue through the powerful statistical technique of multilevel modelling (MLM) \citep{Mandel2009, Loredo2010, Mandel2011, Soiaporn2012}.  This broad class of modelling techniques base prior probabilities on the actual model parameter distribution implied between data sets.  MLM constrains the actual prior distribution by requiring that the distribution derived from the individual measurements match the prior distribution assumed.

In the next section we will introduce the MLM methodology and outline the specific technique used here:  Bayesian Hierarchical Modelling.  In Section 3 we specify our model assumptions.  Finally, we present our results and discuss their implications for characterizing Local Group dSphs.

\section{MultiLevel Modelling}
 
Constraining dark matter halo properties from individual stellar line-of-sight velocity measurements in Local Group dSph spheroidal galaxies is a difficult problem.
Mass constraints from dispersion measurements are riddled with unconstrained degeneracies that affect mass measurements far from the stellar half-light radius \citep{Walker2009a, Wolf2010}.  This causes inferred mass probabilities to be dominated by prior probabilities.  In Bayesian analysis, this is problematic because of the ``degree of belief'' probabilistic interpretation that is
usually assigned to the prior and posterior probabilities \citep{Cox1946}.  In other words, the mass posterior beyond the half-light radius is dominated by the observer's (sometimes arbitrary) prior belief rather than being dominated by data.  
One solution is to apply the strict frequentist interpretation of probability to the prior --  {\em e.g.}, restrict the interpretation of the prior probability density function (PDF) to represent the {\em frequency} of observing an halo property given a sufficiently large galaxy sample.   Within this interpretation, the choice of prior is constrained to match that of the overall galaxy sample.  This causes resultant mass posteriors to be much more stringent.  However, the accuracy of these posteriors is highly dependent on the agreement between the assumed prior probability and the actual galaxy sample distribution. For the Local Group dSphs, the properties of the source galaxy sample is usually inferred from numerical simulations.  Unfortunately, even if these simulations are an adequate description of the underlying distribution, the actual observable distribution will only be a subset of this sample.   Because the subset is determined by astrophysical interactions that currently are not well understood, it is very likely that strict application of numerical simulations, in this regard, will lead to erroneous results.

In this paper we address the issue of prior dominance by applying a multilevel statistical modelling technique to directly constrain the prior probabilities.  
MLM divides the parameters of the dSph galaxies into
various `levels', each with its own set of observables. Starting
with the most basic `lowest' level, the posteriors on the
observables at each level are used as input into subsequent levels.

Starting at the lowest level, the distribution being constrained consists solely of the set of observables, $\mathscr{D}^{(0)} = \{d_i\}$, where the probability of observing a single data point, $d_i$, given a parametrization, $\mathscr{M}^{(0)}_0$, is
\begin{equation}
\mathcal{P}(d_i | \mathscr{M}^{(0)}_0).
\end{equation}
Here, $\mathscr{M}^{(n)}_x$ represents a parametrization that was introduced at the $x$th level and is the $n$th superset consisting of the sets $\{\mathscr{M}^{(n-1)}_{x, i}\}$ (for $n > 0$).
Because there is an {\em actual} distribution of individual data points ($\{d_i\}$), a likelihood function can be defined that can be used to gauge the quality of fit of the assumed set of model parameters, $\mathscr{M}^{(0)}_0$:
\begin{equation}\label{eq:like0}
\mathcal{L}(\mathscr{M}^{(0)}_0 | \mathscr{D}^{(0)}) \equiv \mathcal{P}(\mathscr{D}^{(0)} | \mathscr{M}^{(0)}_0) = \prod_i \mathcal{P}(d_i | \mathscr{M}^{(0)}_0).
\end{equation}
But in order to infer a posterior probability of 
\begin{subequations}\label{eq:joint0}
\begin{eqnarray}
\mathcal{P}(\mathscr{M}^{(0)}_0 | \mathscr{D}^{(0)}) & = & \mathcal{L}(\mathscr{M}^{(0)}_0 | \mathscr{D}^{(0)}) \frac{\mathcal{P}(\mathscr{M}^{(0)}_0)}{\mathcal{P}(\mathscr{D}^{(0)})} \\ & = & \frac{\mathcal{P}(\mathscr{M}^{(0)}_0)}{\mathcal{P}(\mathscr{D}^{(0)})} \prod_i \mathcal{P}(d_i | \mathscr{M}^{(0)}_0).
\end{eqnarray}
\end{subequations}
a prior PDF, $P(\mathscr{M}^{(0)}_0)$, must be supplied.  Since only one set of model parameters, $\mathscr{M}^{(0)}_0$, can be inferred from a single data set, $\mathscr{D}^{(0)}$, we possess little knowledge about the set of $\{\mathscr{M}^{(0)}_0\}$ needed to constrain the prior probability distribution.     
But this can be alleviated if a superset of data is available -- {\em e.g.}, a set of $\mathscr{D}^{(0)}$ ($\mathscr{D}^{(1)} \equiv \{\mathscr{D}^{(0)}_i\}$) from which a set of individual $\mathscr{M}^{(0)}_{0}$'s ($\mathscr{M}^{(1)}_0 \equiv \{\mathscr{M}^{(0)}_{0, i}\}$) can be estimated (note that $\mathscr{D}^{(n)}$ represents the $n$th superset of data consisting of the sets $\{\mathscr{D}^{(n-1)}_i\}$).  
Given this superset of data, $\mathscr{D}^{(1)}$, a new level of modelling can be introduced that simultaneously constrains the full set of $\mathscr{M}^{(1)}_0$.  The only truly free parameters at this level are the ones used to parametrize the prior, $\mathscr{M}^{(0)}_1$.  In this context, this level's likelihood now becomes a function of $\mathscr{M}^{(0)}_1$:
\begin{subequations} \label{eq:like1}
\begin{eqnarray}
\lefteqn{\mathcal{L}(\mathscr{M}^{(0)}_1 | \mathscr{M}^{(1)}_0, \mathscr{D}^{(1)}) \equiv \mathcal{P}(\mathscr{M}^{(1)}_0, \mathscr{D}^{(1)} | \mathscr{M}^{(0)}_1) } \\
& & = \prod_j \mathcal{P}(\mathscr{M}^{(0)}_{0, j}, \mathscr{D}^{(0)}_j | \mathscr{M}^{(0)}_1) \\
& & = \prod_j \mathcal{P}(\mathscr{M}^{(0)}_{0, j} | \mathscr{M}^{(0)}_1) \prod_i \mathcal{P}(d_{j, i} | \mathscr{M}^{(0)}_{0, j})
\end{eqnarray}
\end{subequations}
where $d_{j, i}$ is $i$-th data point of the $j$-th data set.
By adding this additional level of modelling, the prior PDF from the previous level is reinterpreted as a contributing term in the higher level likelihood.  Unfortunately, the probability distribution,  $\mathcal{P}(\mathscr{M}^{(0)}_1)$, is not specified for the same reason that $\mathcal{P}(\mathscr{M}^{(0)}_0)$ could not be constrained in the previous level:  at this level, only one $\mathscr{M}^{(0)}_1$ can be inferred.  This necessitates the assumption of an unrestrained prior PDF to determine the joint posterior probability distribution, 
\begin{equation}
\mathcal{P}(\mathscr{M}^{(0)}_1, \mathscr{M}^{(1)}_0 | \mathscr{D}^{(1)})  =  \mathcal{L}(\mathscr{M}^{(0)}_1 | \mathscr{M}^{(1)}_0, \mathscr{D}^{(1)}) \frac{\mathcal{P}(\mathscr{M}^{(0)}_1)}{\mathcal{P}(\mathscr{D}^{(1)})}.
\end{equation}
Thus, in this modelling methodology, information contained within the internal distribution of each data set, as well as the distribution of data between each data set, is concurrently used to constrain the combined probability distributions $\mathcal{P}(d_i | \mathscr{M}^{(0)}_0)$ as well as the prior distribution, $\mathcal{P}(\mathscr{M}^{(0)}_i | \mathscr{M}^{(0)}_1)$. 

Of course, given a set of $\mathscr{D}^{(1)}$ ($\mathscr{D}^{(2)} = \{\mathscr{D}_i^{(1)}\}$), $P(\mathscr{M}^{(0)}_1)$ can be constrained by the application of an additional level of modelling.  In this manner, MLM can be used recursively to constrain newly introduced prior probabilities from previous iterations -- provided the appropriate superset of data ($\mathscr{D}^{(n)}$) is available.  Used recursively, the $n$-th implementation ($n$-th level) has a likelihood function of
\begin{subequations} \label{eq:like1}
\begin{eqnarray}
\lefteqn{\mathcal{L}(\mathscr{M}^{(0)}_n | \mathscr{M}^{(n)}_0, \mathscr{M}^{(n-1)}_1, \dots, \mathscr{M}^{(1)}_{n-1}, \mathscr{D}^{(n)})} \\
& & \equiv \mathcal{P}(\mathscr{M}^{(n)}_0, \mathscr{M}^{(n-1)}_1, \dots, \mathscr{M}^{(1)}_{n-1}, \mathscr{D}^{(n)} | \mathscr{M}^{(0)}_n) \\
&  & =  \prod_i \mathcal{P}(\mathscr{M}^{(0)}_{n-1, i} | \mathscr{M}^{(0)}_n) \mathcal{L}(\mathscr{M}^{(0)}_{n-1, i} | \dots, \mathscr{D}^{(n-1)}_i)
\end{eqnarray}
\end{subequations}
where $\mathscr{M}^{(n)}_x \equiv \{\mathscr{M}^{(n-1)}_{x, i}\}$ and $\mathscr{D}^{(n)} \equiv \{\mathscr{D}^{(n-1)}_i\}$.

Regardless of the number of levels applied, the prior introduced at the top-level is still completely unconstrained.  Thus, to utilize the complete posterior distribution in a statistically consistent manner, either the top-level prior must be inferred from external information or the interpretation of probability must be expanded to include the Bayesian `degree of belief' probabilistic interpretation.  Although logically consistent, the `degree of belief' interpretation introduces subjectivity that is unsettling to some scientists.  And, the applicability or accessibility of prior information can make the former methodology difficult to implement.  However, the  subjectivity of the top-level prior has only an indirect, and thus mitigated, effect on lower-level posteriors.  This is because this prior affects lower-level posteriors only indirectly through lower-level priors.  The strict frequentist interpretation applied to the likelihoods at various levels and their associated lower-level priors ensure that these lower-level priors are constrained by the intrinsic distribution of the data. This, thereby, mitigates the effect of the top-level prior assumptions.

In this paper, we utilize the full top-level posterior
\begin{equation}
\mathcal{P}(\mathscr{M}^{(0)}_n, \dots | \mathscr{D}^{(n)})  =  \mathcal{L}(\mathscr{M}^{(0)}_n | \dots, \mathscr{D}^{(n)}) \frac{\mathcal{P}(\mathscr{M}^{(0)}_n)}{\mathcal{P}(\mathscr{D}^{(n)})}.
\end{equation}
to obtain model parameter uncertainties.  Although the rather arbitrary Bayesian `degree of belief' probabilistic interpretation is assigned to the top-level prior, these so-called `Bayesian hierarchical models' \citep{Loredo2010} provide a straight forward and statistically consistent methodology to apply MLM.  Another option would be to use `Bayesian empirical modeling' which (ironically) avoids the introduction of the Bayesian probabilistic interpretation by solely utilizing the top-level maximum likelihood solution \citep{Berger1985, Petrone2012}.  But, these methods are usually not as conservative (as if the full posterior were used) because they do not explore the full top-level parameter space.

\section{model assumptions}

In this paper we assume a two level model. 
The bottom-level describes the astrophysical properties of each individual dSph and its underlining dark matter potential, whereas the top-level details the overall distribution of halo properties.  For the bottom-level, the total set of observables are the line-of-sight velocities, metallicites, and positions of individual stars in the galaxy, as well as the total galaxy luminosity.  The total model parameter set is composed of the stellar profile, dark matter profile, and stellar velocity anisotropy parameters.  Normally, a likelihood would be created that models the intrinsic dispersion of each galaxy via the Jeans equation.  To save computational time, we utilize the approximation that the Jeans equation only constrains the enclosed mass at the (3D) half-light radius.  Specifically, we utilize the relationship between the mass at the half-light radius ($M(r_{1/2})$) and total measured velocity dispersion ($\sigma_{\textrm{tot}}^2$) derived in \citet{Wolf2010}:
\begin{equation}
M(r_{1/2}) = 3 \sigma_{\textrm{tot}}^2/G.
\end{equation}
Here, it is the mass profile ($M(r)$) that is parametrized by the lower-level parameters ($\mathscr{M}^{(0)}_0$).  This approximation has been found to be very accurate because the degeneracy between the enclosed mass and the stellar velocity anisotropy is drastically minimized at the stellar half-light radius causing the mass posterior at this point to be data dominated \citep{Walker2009a, Wolf2010}.  This approximation has the additional benefit of removing any potential biases caused by the assumed form of the velocity anisotropy profile.  Using this approximation, the bottom-level data set is now composed of the mass enclosed within the half-light radius ($M_{1/2}$, or equivalently, the measured velocity dispersion), the measured half-light radius ($r_{1/2}$), the total luminosity ($L$), and their associated errors ($\epsilon_{M_{1/2}}$, $\epsilon_{r_{1/2}}$, and $\epsilon_{L}$) -- thus, 
\begin{equation} \label{eq:dwarfdata}
\mathscr{D}^{(0)} = \{M_{1/2}^{\textrm{(obs)}}, \epsilon_{M_{1/2}}, r_{1/2}^{\textrm{(obs)}}, \epsilon_{r_{1/2}}, L^{\textrm{(obs)}}, \epsilon_{L}\}
\end{equation}
\citep[and references within]{Wolf2010, McConnachie2012}.  To better match the actual distributions, we model $M_{1/2}$ and $L$ as lognormal distributions, and $r_{1/2}$ as a normal distribution:
\begin{eqnarray}
\lefteqn{\mathcal{L}(\mathscr{M}^{(0)}_0 | \mathscr{D}^{(0)}) = \mathcal{N}(\textrm{log}(M(r_{1/2})), \epsilon_{M_{1/2}} | \rm{log}(M_{1/2}^{\textrm{(obs)}})) } \nonumber \\
& & \times \mathcal{N}(\textrm{log}(L), \epsilon_{L} | \textrm{log}(L^{\textrm{(obs)}})) 
\times \mathcal{N}(r_{1/2}, \epsilon_{r_{1/2}} | r_{1/2}^{\textrm{(obs)}}) 
\end{eqnarray}
Here, $\mathcal{N}(\mu, \sigma | x)$ denotes a normal distribution in $x$ defined by:
\begin{equation}
\mathcal{N}(\mu, \sigma | x) \equiv \mathcal{P}_{\textrm{norm}}(x | \mu, \sigma) = \frac{1}{\sqrt{2\mathrm{\pi}}\sigma}\mathrm{exp}\left[-\frac{(x-\mu)^2}{2\sigma^2}\right].
\end{equation}
As $r_{1/2}$ has little implication for the underlying dark matter theory, we assume a non-informative uniform Jeffery's prior and marginalize over all possible $r_{1/2}$ values.  The remaining bottom-level prior then consists of $L$ and the parameters that model the enclosed mass ($M(r)$).  Since the enclosed mass is dominated by the dark matter contribution, its properties represent the dark matter distribution within the galaxy.  Various dark matter halo properties have been found to have tight correlations in the $\Lambda$CDM paradigm.  Particularly, relationships between the halo concentration and mass \citep{Diemand2007, Neto07, Strigari07-redef, Springel2008} as well as between the mass and luminosity \citep{Koposov2009, Busha2010, Font2011, Rashkov2012}.  We use these correlations as motivation for the selection of the functional form of the prior.  Using the profile-independent quantities of the maximum circular velocity ($v_{\textrm{max}}$) and radius corresponding to this velocity ($r_{\textrm{max}}$) as proxies for halo concentration and mass respectively, the bottom-level prior becomes
\begin{eqnarray} \label{eq:botprior}
\lefteqn{{\mathcal{P}(\log(v_{\textrm{max}}), \log(r_{\textrm{max}}), \log(L))}}\nonumber\\& & = \mathcal{P}(\log(r_{\textrm{max}}) | \log(v_{\textrm{max}}))  \mathcal{P} (\log(v_{\textrm{max}}) | \log(L))\nonumber \\ & & \times \mathcal{P}(\log(L)).
\end{eqnarray}
The first term in the RHS of Equation \ref{eq:botprior} represents the $r_{\max}$-$v_{\max}$ relation.  Simulations show that this relation closely resembles a linear relationship between  log($r_{\max}$) and log($v_{\max}$) with an intrinsic dispersion \citep{Diemand2007, Strigari07-redef, Springel2008}:
\begin{equation}
\mathcal{P}(\log(r_{\max}) | \log(v_{\max})) \approx \mathcal{N}(\alpha_{rv}\log(v_{\max}) + \beta_{rv}, \sigma_{rv})
\end{equation}
Here, we keep this functional form, but do not infer the model parameters from simulations.  Rather, we constrain these parameters using the next level's likelihood.  For the mass--luminosity relation encapsulated in $\mathcal{P}(\log(v_{\textrm{max}}) | \log(L))$, we again assume a log--log relationship with some intrinsic dispersion:
\begin{equation}
\mathcal{P}(\log(v_{\max}) | \log(L)) \approx \mathcal{N}(\alpha_{vl}\log(L) + \beta_{vl}, \sigma_{vl})
\end{equation}
Our choice is motivated by observations implying that this relationship is flat \citep{Strigari2007, Strigari2008} and by the functional form of mass distributions in simulations that account for completeness and re-ionization effects \citep{Koposov2009, Busha2010, Font2011, Rashkov2012}.  Since these probability distributions depend only conditionally on luminosity, the functional forms of these probability distributions are not expected to be influenced by observational bias in luminosity and therefore describe the underlying dark matter distribution properties.  This is not true for the luminosity function, $\mathcal{P}(\log(L))$, which is affected significantly by observational bias.  Thus, the functional form of $\mathcal{P}(\log(L))$ is expected to be a complicated convolution of this bias and the actual underlying luminosity function.  However, this effect tends to flatten the $\mathcal{P}(\log(L))$ at faint luminosities \citep{Koposov2008}.  Here, we assume the simplified functional form:
\begin{equation}
\mathcal{P}(\log(L)) \propto L^{\alpha_l}.
\end{equation}
For the underlying dark matter density profiles ($\rho(r)$) we consider the four models listed in Table \ref{tab:evidence}:  NFW, Cored NFW, Burkert, and Einasto defined as
\begin{equation}
\rho(r) = \left\{
\begin{array}{cc}
\frac{\rho_{\textrm{S}} r_{\mathrm{S}}^3}{r(r_{\mathrm{S}} + r)^2} & \text{NFW} \\
\frac{\rho_{\mathrm{S}} r_{\mathrm{S}}^3}{r_c r_{\mathrm{S}}^2+ r(r_{\mathrm{S}} + r)^2} & \text{Cored NFW} \\
\frac{\rho_{\mathrm{S}} r_{\mathrm{S}}^3}{(r_{\mathrm{S}} + r)(r_{\mathrm{S}}^2 + r^2)} & \text{Burkert} \\
\rho_{\mathrm{S}} \exp\left[-2n\left(\left(\tfrac{r}{r_{\mathrm{S}}}\right)^{1/n} - 1\right)\right] & \text{Einasto}
\end{array} \right.
\end{equation}
where $r_S$ and $\rho_S$ are derived from the profile-independent quantities $v_{\textrm{max}}$ and $r_{\textrm{max}}$.

Therefore, the complete set that parametrize the bottom-level likelihood is 
\begin{equation}
\mathscr{M}^{(0)}_0 = \{v_{\textrm{max}}, r_{\textrm{max}}, L\}, 
\end{equation}
whereas the complete sets that parameterizes the bottom-level prior for each density profile model are 
\begin{equation}
\mathscr{M}^{(0)}_1 = \left\{
\begin{array}{cc}
	\left\{
	\begin{array}{cccc} 	\alpha_{rv}, & \beta_{rv}, & \sigma_{rv}, & \\
						\alpha_{vl}, & \beta_{vl}, & \sigma_{vl}, & \alpha_{l}
	\end{array}
	\right\} & \begin{array}{c}\text{NFW/} \\ \text{Burkert}\end{array} \\
\left\{
	\begin{array}{cccc} 	\alpha_{rv}, & \beta_{rv}, & \sigma_{rv}, & r_{\mathrm{c}}/r_{\mathrm{S}}, \\
						\alpha_{vl}, & \beta_{vl}, & \sigma_{vl}, & \alpha_{l}
	\end{array}
	\right\} & \text{Cored NFW} \\
\left\{
	\begin{array}{cccc} 	\alpha_{rv}, & \beta_{rv}, & \sigma_{rv}, & n, \\
						\alpha_{vl}, & \beta_{vl}, & \sigma_{vl}, & \alpha_{l}
	\end{array}
	\right\} & \text{Einasto.}
\end{array} \right.
\end{equation}
These sets are in turn constrained by the top-level likelihood (Equation \ref{eq:like1}).  The top-level superset of data ($\mathscr{D}^{(1)}$) is composed of individual data sets (Equation \ref{eq:dwarfdata}) of the 20 dSphs listed in Table \ref{tab:params}.  Standard non-informative priors were assumed for the top-level priors:  uniform in $\alpha_x$, $\beta_x$, and $\log(\sigma_x)$ whose prior ranges are given in Table \ref{tab:limits}.  
We also assume uniform priors for the cored NFW scaled core radius ($r_{\mathrm{c}}/r_{\mathrm{S}}$) and Einasto index ($n$) limited to the range $0 < r_{\mathrm{c}}/r_{\mathrm{S}} < 1$ and $0.5 < n < 10$ respectively.  The allowed range for the luminosities is also marginalized over with the exception of the Einasto model whose luminosity range, because of computational complications, was set to be $2.0 < \log_{10}(L) < 8.0$.  For the other models, the lower limit to $\log_{10}(L)$ was allowed to vary between $0$ and $5$ whereas the upper limit was allowed to vary between $5$ and $10$. 
However, in next section we show that the form of the luminosity function, such as the slope or allowed luminosity range, affects neither the final lower-level posteriors nor the higher-level parameters of interest.
In total, this model contains
67--68 parameters ($20 \times \mathscr{M}^{(0)}_0 + \mathscr{M}^{(0)}_1$), though each galaxies'
$L_i$ parameter is analytically integrated to reduce the number of
parameters by 20.  To explore this parameter space, we employ a metropolis nested sampling technique \citep{Skilling2004, Brewer2010} obtaining approximately 500 000 sample points per run.  In the next section we show that these results not only diminish systematic uncertainties due to the degeneracy between the enclosed mass and the velocity anisotropy, but also directly constrain the properties of the Milky Way dSph satellite distribution.

\section{Results}
\label{section:results}
%
\renewcommand{\thefootnote}{\fnsymbol{footnote}}
\begin{table}
\caption{\label{tab:evidence}Summary of the Bayes factors.}
\scriptsize
\begin{center}
\begin{tabular}{@{}cccccccl}
\noalign{\hrule height 1pt}
\vspace{1pt}
Model & log$_{e}$(Bayes factor\footnotemark[1]) \\ 
\hline
\vspace{1pt}
NFW & $0$ \\ 
Burkert & $1.03 \pm 0.36$ \\ 
Cored NFW & $2.19 \pm 0.34$ \\ 
Einasto & $-16.42 \pm 0.35$ 
\vspace{1pt}\\
\noalign{\hrule height 1pt}
\end{tabular}
\end{center}
\begin{flushleft}
\footnotemark[1]{Here the Bayes factor is the ratio of the evidences ($\mathscr{E}$) relative to the evidence of the NFW model ($\mathscr{E}_{\textrm{model}}/\mathscr{E}_{\textrm{NFW}}$).}
\end{flushleft}
\end{table}

\begin{figure*}
\begin{center}
\includegraphics[width=0.96\hsize]{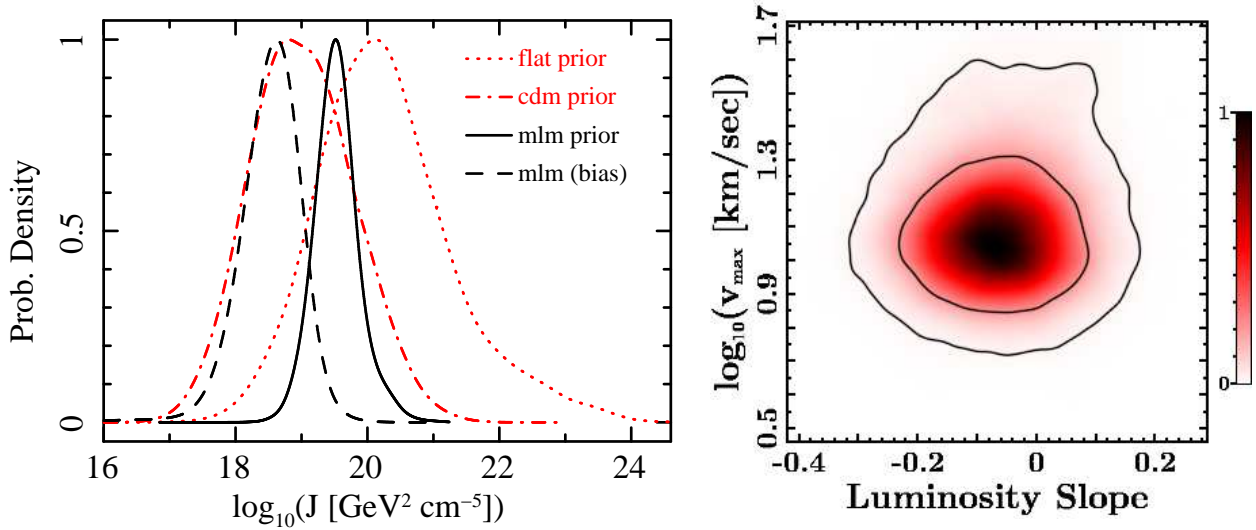}
\caption {
Left: this figure illustrates the ability of this methodology to limit prior dominance on lower-level posteriors.  Compared are the $J$ factors posteriors using MLM to that using assumed lower-level priors.  Plotted are the $J$ factor assuming a prior that resembles the distribution predicted by CDM simulations (CDM prior, dotted red line), priors uniform in $\log(r_{\textrm{max}})$ and $\log(v_{\textrm{max}})$ (dash-dotted red line), and two posteriors that employ the MLM methodology presented in this paper.  Because simulations predict a large amount of low mass sub-haloes, the CDM prior assumptions bias the $J$ posterior artificially to lower values.  The result is a posterior that is significantly lower than if uniform priors where assumed.  For comparison, the posteriors of two MLM runs are also plotted:  one result assumes the usual non-informative priors (black solid line) and the other result drastically biases our posterior results to artificially low concentrations.  Not only are these two posteriors more robust to top-level assumptions, but the resulting posteriors are better constrained.
Right: shown is the joint posterior of the maximum circular velocity ($v_{\textrm{max}}$) and the slope of the luminosity function.  Unsurprisingly there is no apparent correlation between the shape of the luminosity function (e.g. the slope) and lower-level posteriors.  This is because of, to the lower-level posteriors, the effective prior that influences the result is the prior integrated over all possible luminosities (see Section \ref{section:results}, especially Equation \ref{eq:effprior}).
\label{fig:prioreffects}
}
\end{center} 
\end{figure*}
The previously mentioned anisotropy-mass degeneracy that is intrinsic in Jeans modelling makes mass modelling inherently dominated by prior assumptions.  Because MLM constrains the overall distribution (and thus the lower-level priors), this approach can significantly reduce the effect of this degeneracy.  Fig. \ref{fig:prioreffects} (left-hand panel) illustrates the advantage of this methodology in limiting prior dominance on lower-level posteriors. This figure shows the impact of varying prior assumptions on the astrophysical contribution of the dark matter annihilation flux, the $J$ factor \citep{Strigari2007, Martinez2009, Fermi2010, Llena-Garde2011, Charbonnier2011, Fermi2011}.  Shown is one of the most prior dominated galaxies:  Segue 1.  Compared are the solutions of assuming two priors, one non-informative and the other biasing the solution to low masses, with and without the use of hierarchical modelling.  Varying prior assumptions without the use of hierarchical priors can alter the $J$ factor posteriors by more than an order of magnitude.  However, this effect is minimized when hierarchical priors are used.

\begin{figure*}
\begin{center}
\rotatebox{0}{\includegraphics[height=0.32\hsize]{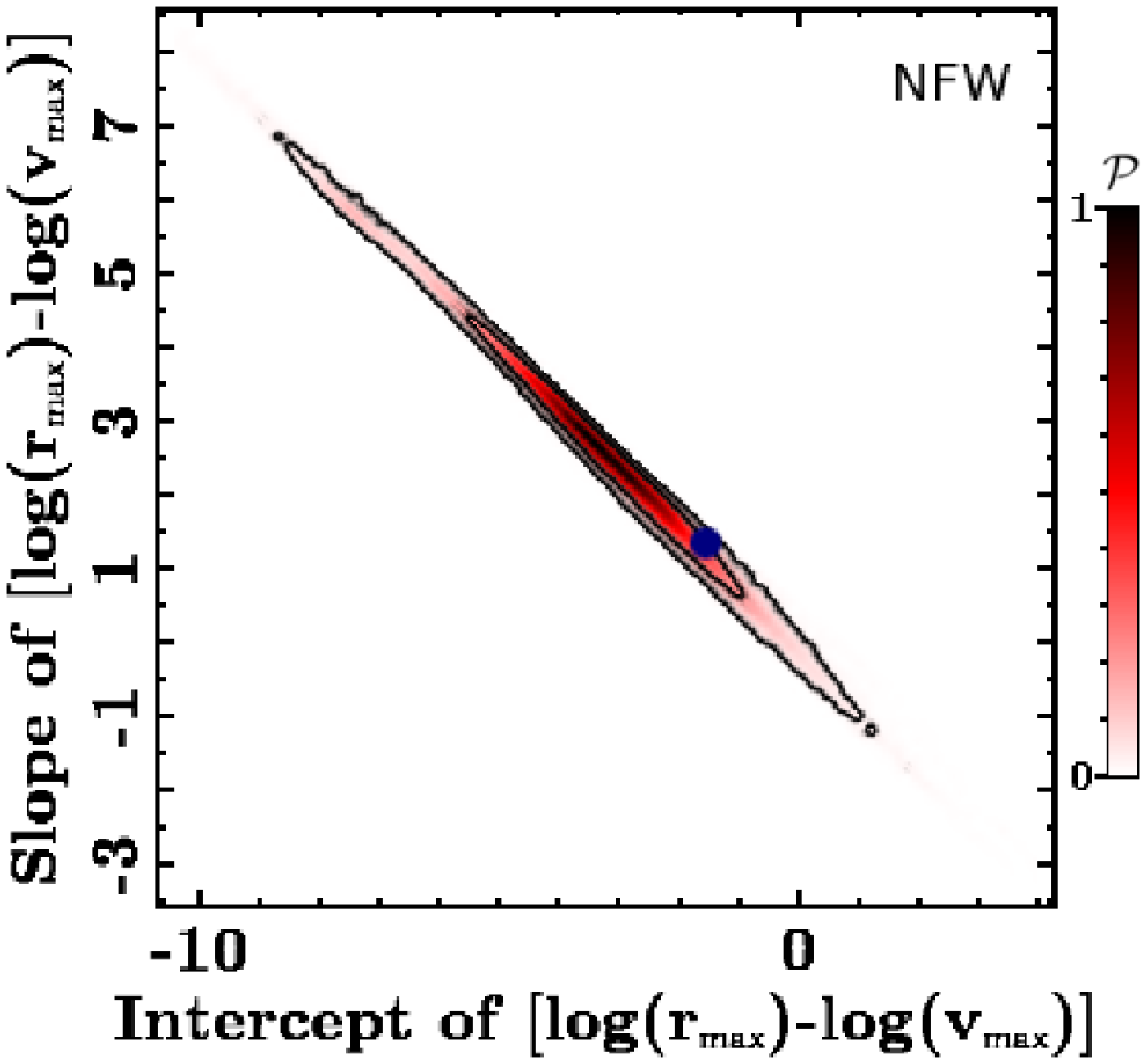}}
\rotatebox{0}{\includegraphics[height=0.32\hsize]{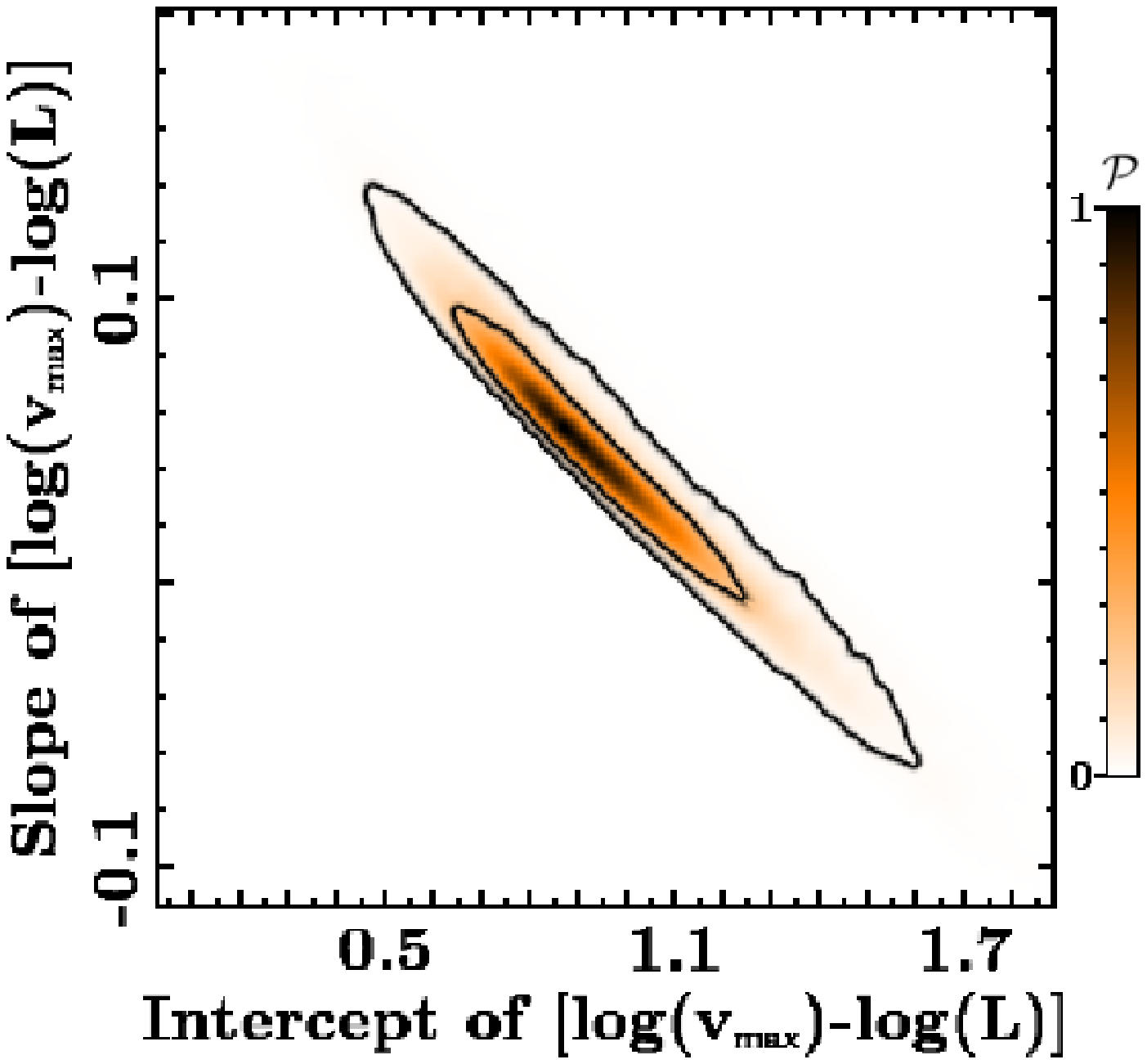}}
\rotatebox{0}{\includegraphics[height=0.3\hsize]{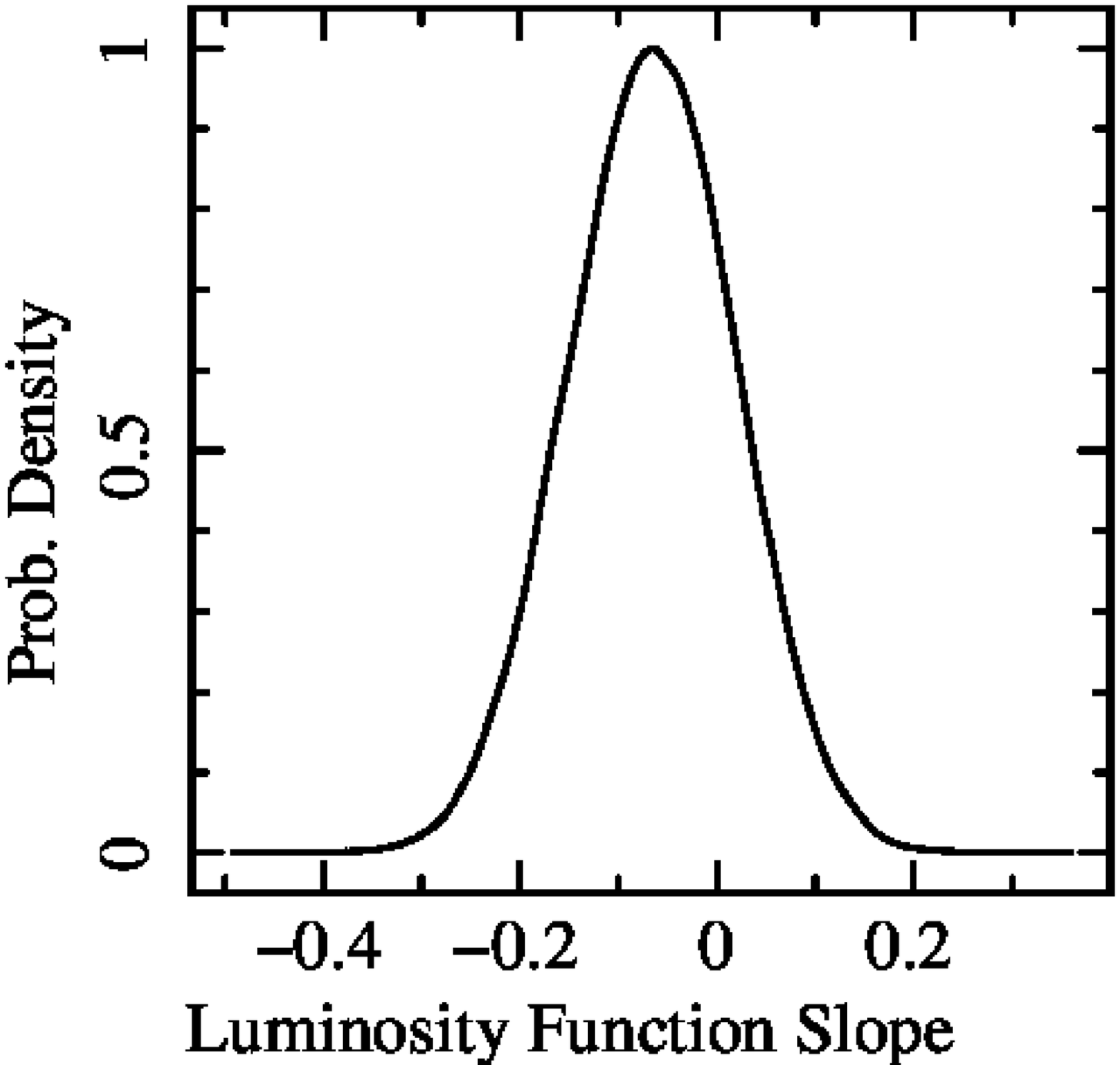}}
\rotatebox{0}{\includegraphics[height=0.32\hsize]{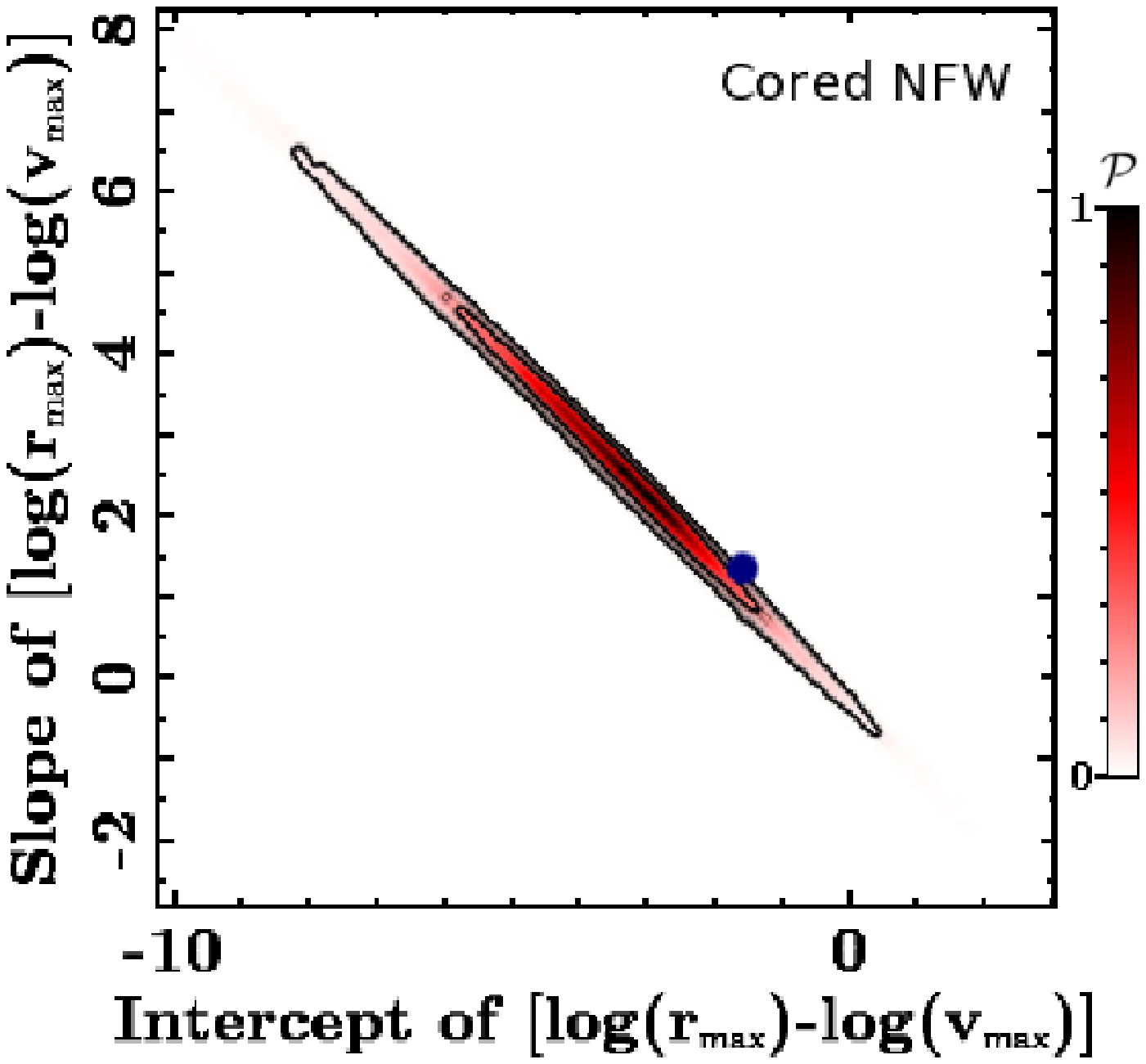}}
\rotatebox{0}{\includegraphics[height=0.32\hsize]{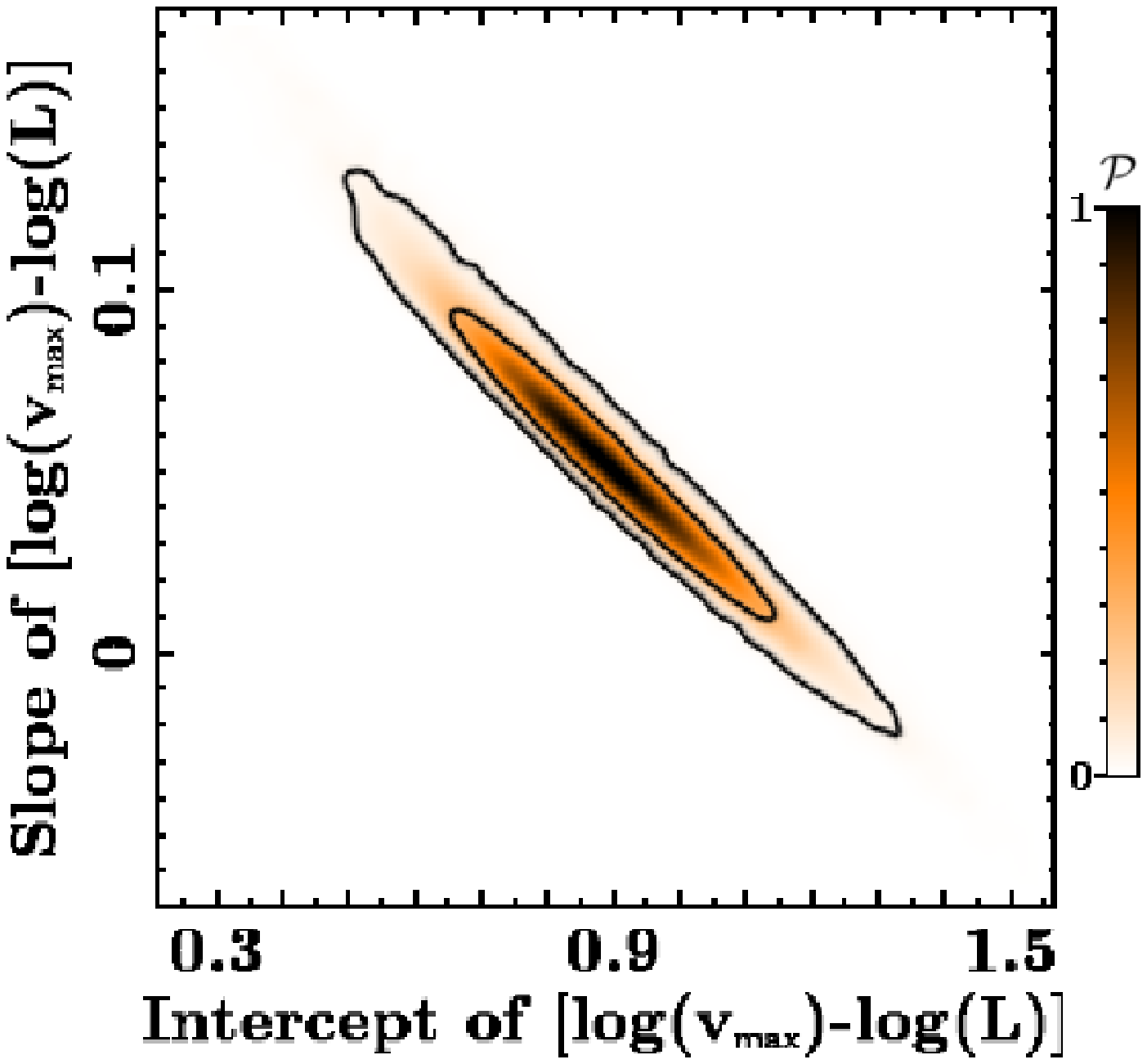}}
\rotatebox{0}{\includegraphics[height=0.3\hsize]{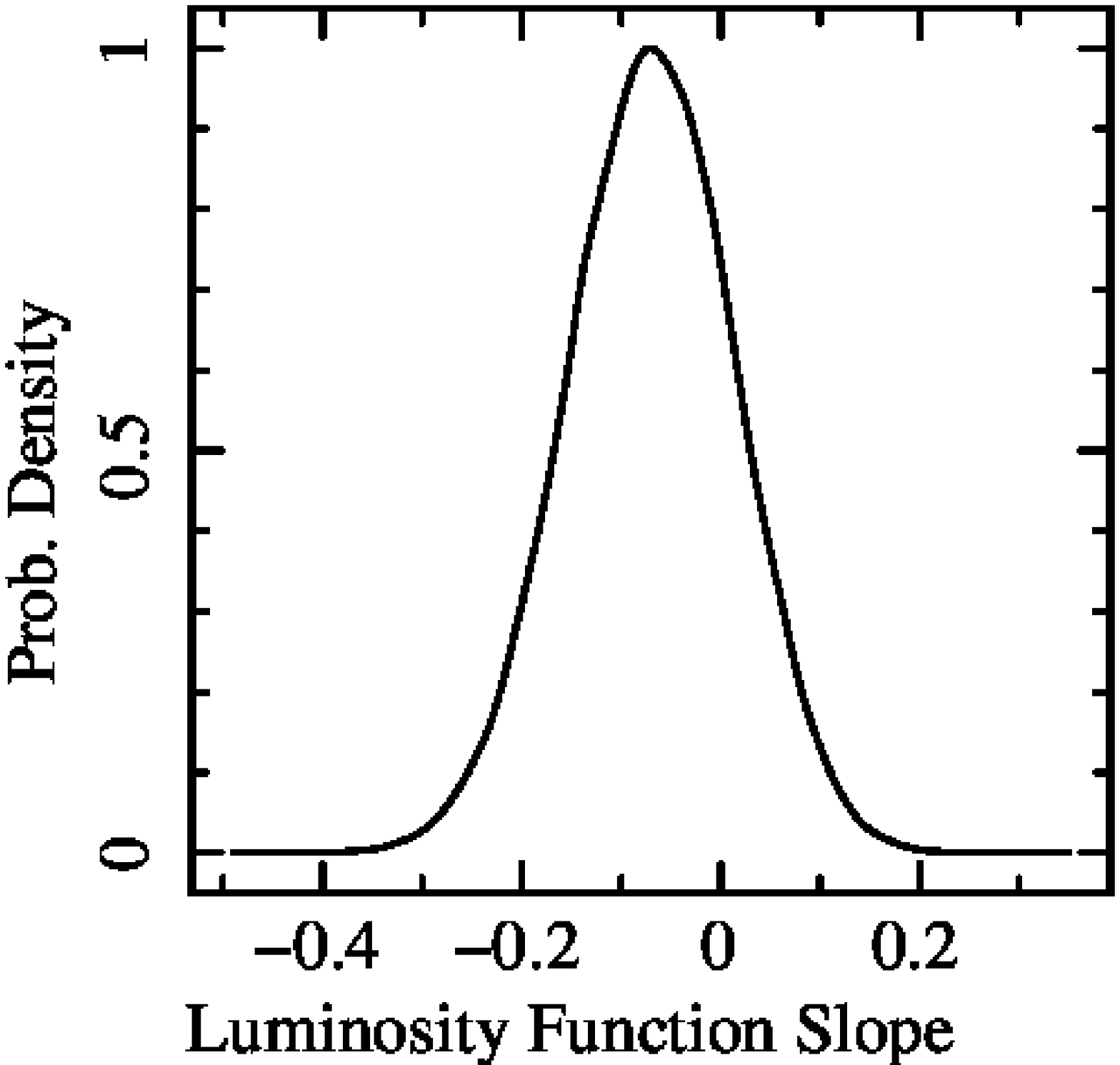}}
\rotatebox{0}{\includegraphics[height=0.32\hsize]{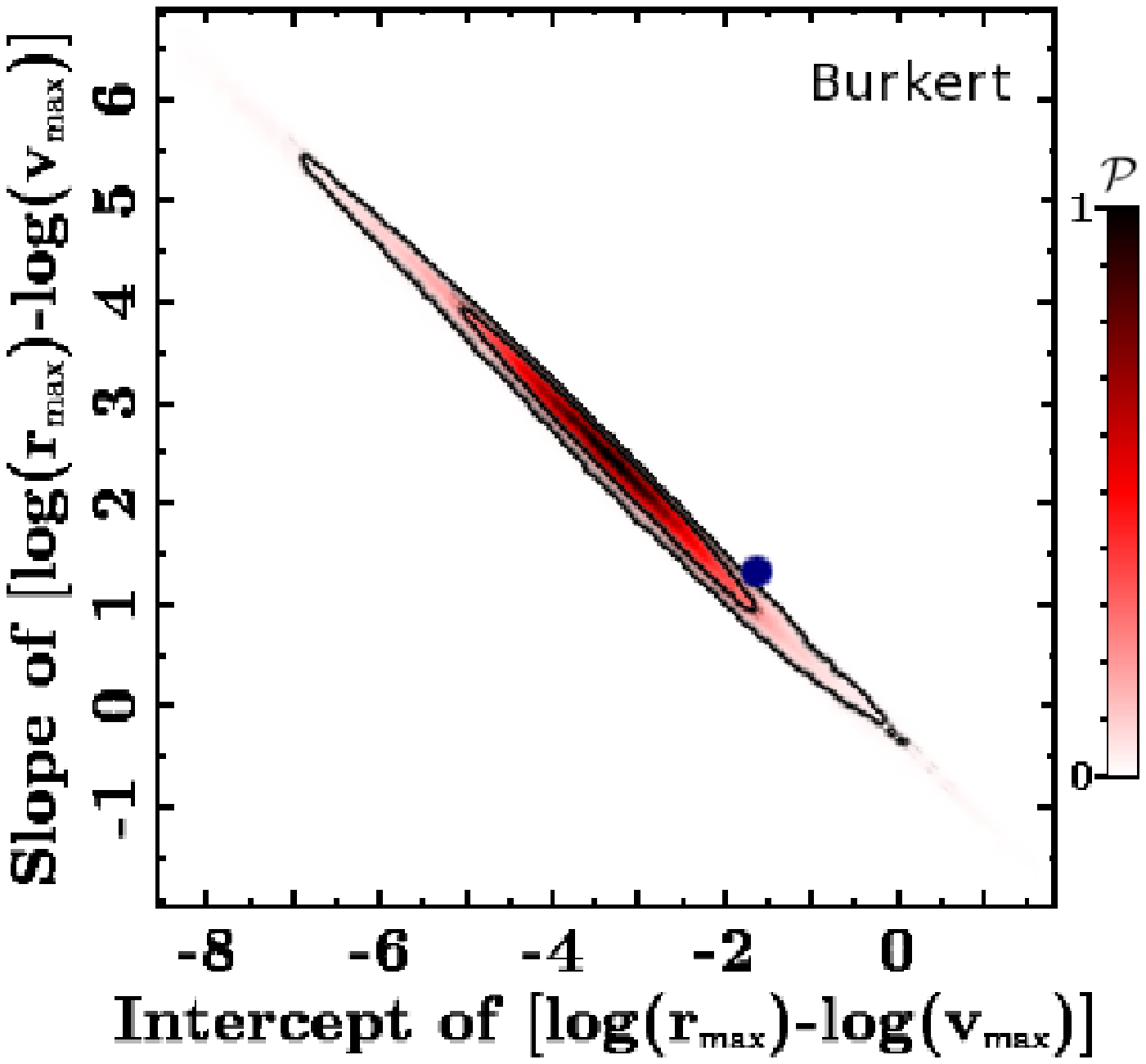}}
\rotatebox{0}{\includegraphics[height=0.32\hsize]{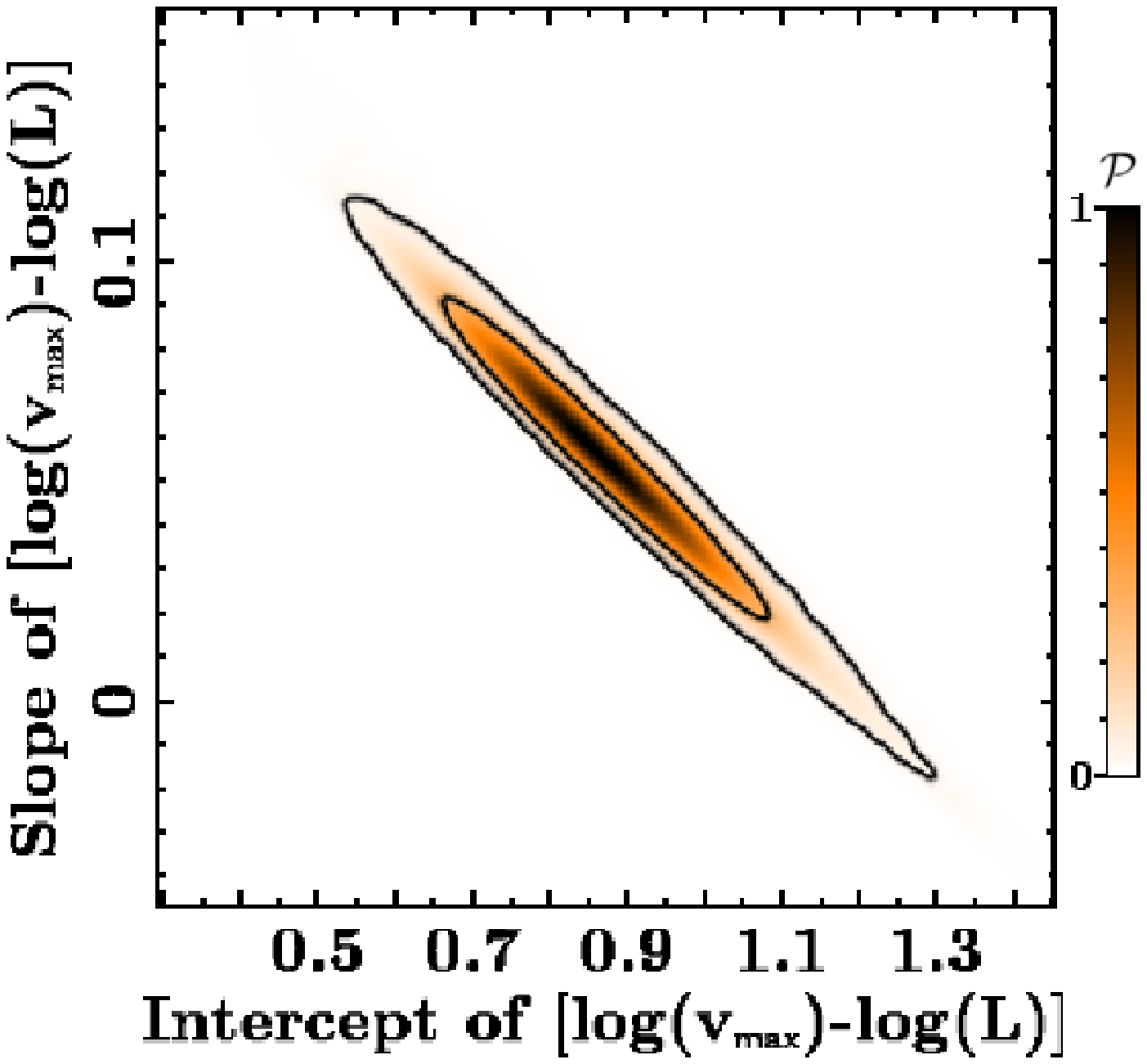}}
\rotatebox{0}{\includegraphics[height=0.3\hsize]{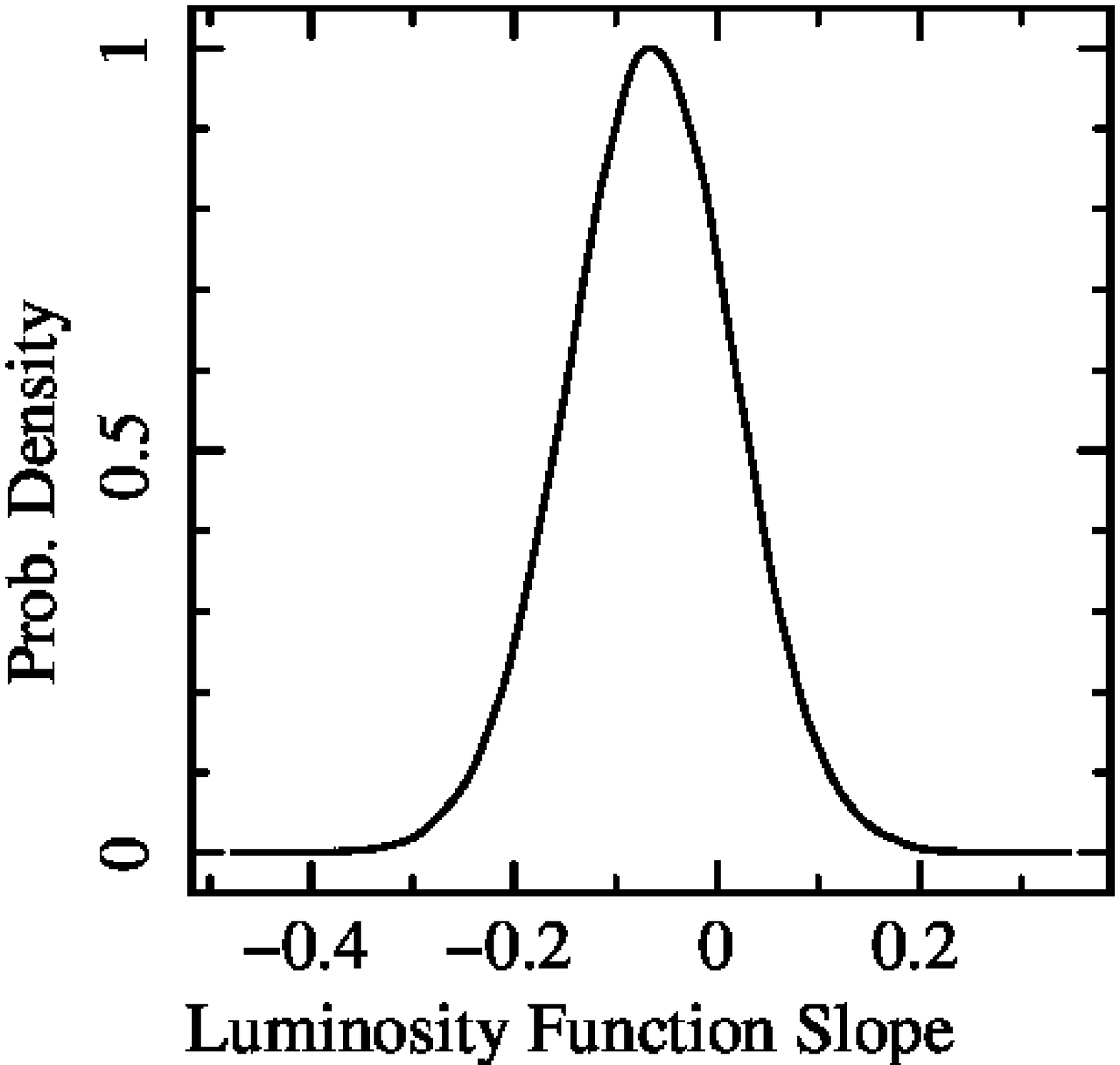}}
\rotatebox{0}{\includegraphics[height=0.32\hsize]{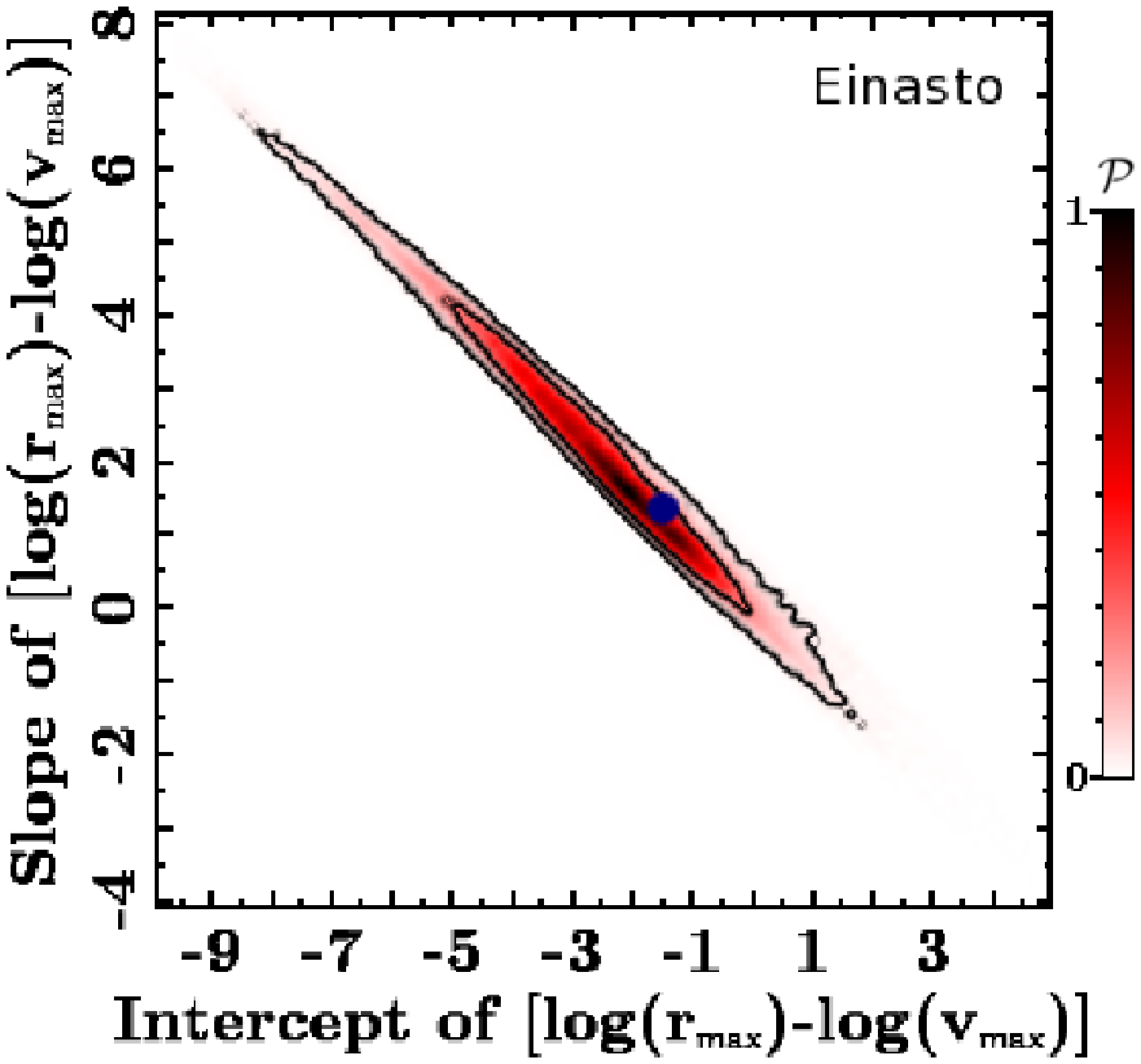}}
\rotatebox{0}{\includegraphics[height=0.32\hsize]{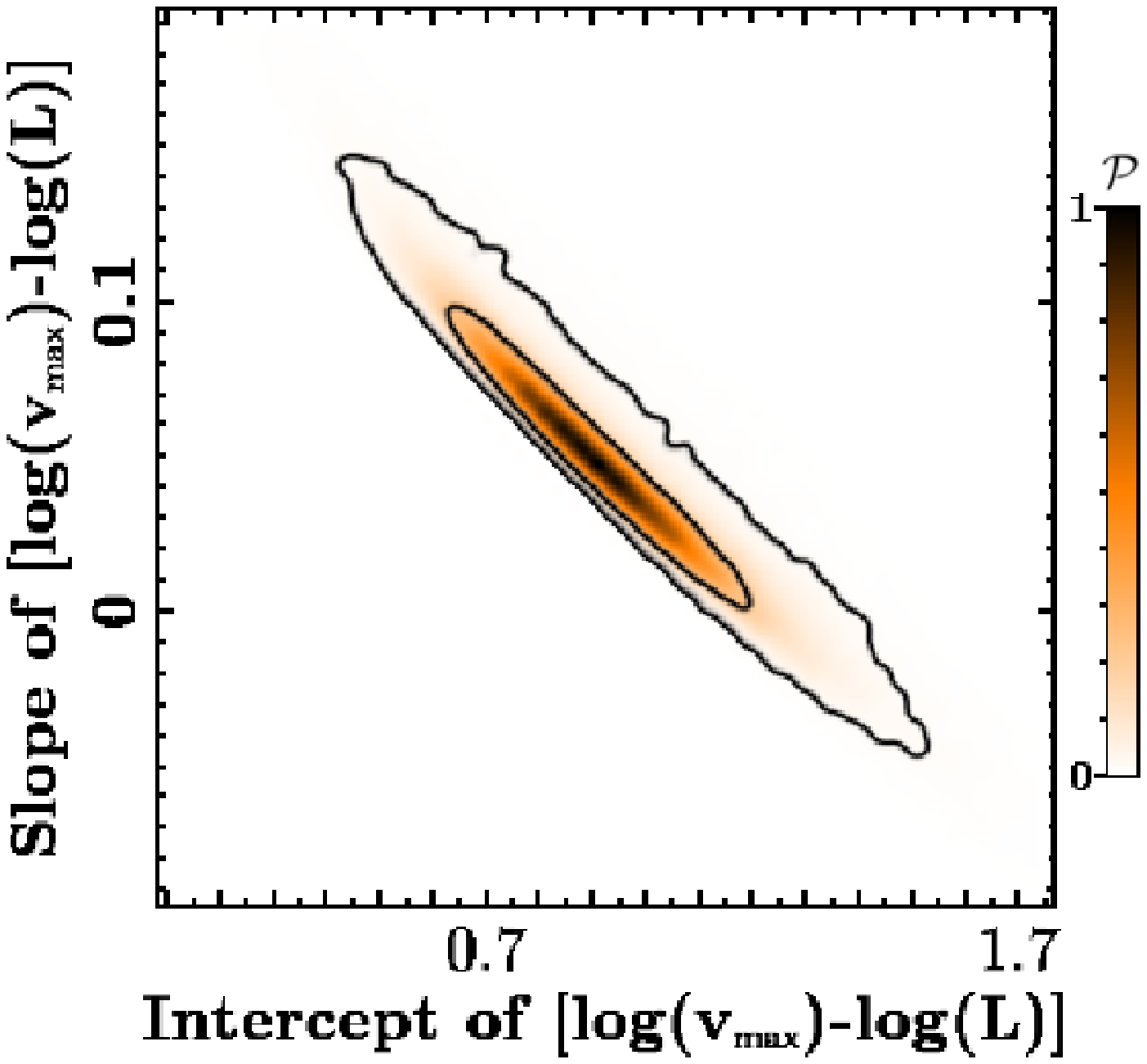}}
\rotatebox{0}{\includegraphics[height=0.3\hsize]{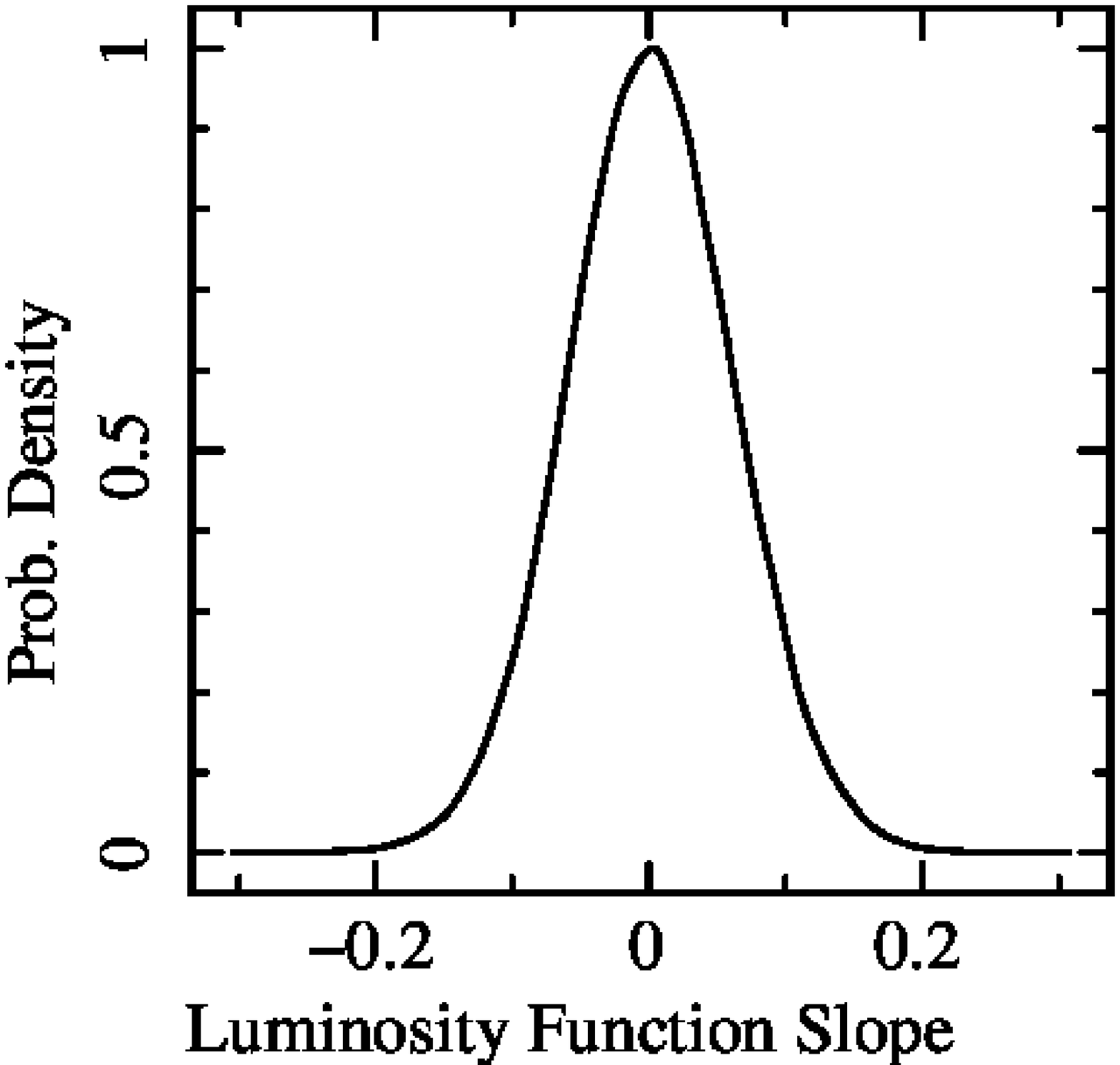}}
\caption {
This figure shows the unrestrained top-level parameter posteriors ($\mathcal{P}$).
Shown in each column, from left to right, are the joint posteriors $\alpha_{rv}$--$\beta_{rv}$ and $\alpha_{lv}$--$\beta_{lv}$, as well as the posterior for $\alpha_{l}$.  From top to bottom, each row contains the posteriors assuming the NFW, cored NFW, Burkert, and Einasto models.  
The blue dot represents the $\alpha_{rv}$ and $\beta_{rv}$ values predicted by simulations \citep{Diemand2007, Strigari07-redef, Springel2008}.  The solid lines represent the 68\% and 95\% credible levels (all $\log$s are base $10$).
\label{fig:toplvlparams}
}
\end{center} 
\end{figure*}

The constraint that MLM provides to prior distributions is not only useful in limiting the effect of prior assumptions on lower-level posteriors, but is also useful in inferring the distribution of galaxy properties.  For example, the distribution of the mass within 300pc as well as the relation between $v_{\textrm{max}}$ and $r_{\textrm{max}}$ have specific simulated predictions.   In the first column in Fig. \ref{fig:toplvlparams}, we show the joint $\alpha_{rv}$--$\beta_{rv}$ posteriors for the four models considered here.  These posteriors are in excellent agreement with CDM simulations \citep[shown by the blue mark]{Diemand2007, Springel2008, Strigari07-redef}.  Unfortunately, while
the effect of the degeneracy between the enclosed mass and velocity anisotropy is minimized at the lower-level posteriors,  this degeneracy manifests itself through an `$\alpha_{rv}$--$\beta_{rv}$ degeneracy'.  This is a consequence of the Milky Way dwarfs approximately sharing the same scale.  But even so, the spread in scale radii is enough to partly constrain this degeneracy.  This effect also manifests itself in the $\alpha_{vl}$--$\beta_{vl}$ joint posteriors (second column), but to a much lesser extent.  It is actually the constraint of the $\mathcal{P}(\log(v_{\max}) | \log(L))$ prior that gives this method this drastic improvement.  This is to be expected since we are including information (luminosity) that was not included in previous studies.  In short, we are constraining our selection of galaxies to be part of a physical galaxy distribution that is dependent on luminosity -- thus, constraints on this underlying distribution will naturally lead to more robust individual galaxy property determination.  The last column in Fig. \ref{fig:toplvlparams} shows the posterior of the slope of the luminosity function.  Out of the lower-level priors constrained, the luminosity function is the only prior directly affected by observation bias.  This could very well affect the form of the luminosity function, even though this effect may still produce a luminosity function close to a power law \citep[see][]{Tollerud2008}.  Fortunately, even though oversimplification of the observation bias may lead to systematic biases on $\alpha_{vl}$ and $\beta_{vl}$, there is only a minimal, if any, influence on the individual lower-level posteriors or on the top-level posteriors  $\alpha_{rv}$ and $\beta_{rv}$.  The reason for this is that the parameters of interest do not explicitly depend on the luminosity in the lower-level likelihoods.  Thus, we may replace the joint $v_{\textrm{max}}$--$L$ prior with
\begin{eqnarray}
\label{eq:effprior}
\lefteqn{\mathcal{P}^{\mathrm{eff}}(\log(v_{\max}) | \log(L^{\mathrm{(obs)}})) \propto \int d(\log(L)) \mathcal{P}(\log(L)) }  \nonumber\\
& & \times \mathcal{P}(\log(L^{\mathrm{(obs)}})) | \log(L))  \mathcal{P}(\log(v_{\max}) | \log(L))\nonumber
\end{eqnarray}
where it is $\mathcal{P}^{\mathrm{eff}}(\log(v_{\max}) | \log(L_{\mathrm{meas}}))$ that is effectively being constrained.  Consequently, if the lower-level parameters are the only parameters of interest, then one may completely forgo specifying the luminosity function and model $\mathcal{P}^{\mathrm{eff}}(\log(v_{\max}) | \log(L_{\mathrm{meas}}))$ directly.  This effect can be seen in Fig. \ref{fig:prioreffects} (right-hand panel) where changes in the luminosities function's shape does not affect the lower-level posteriors.  However in this study, we use the full joint prior in order to produce $v_{\textrm{max}}$--$L$ posteriors that can be consistently compared to future simulations.

\begin{figure*}
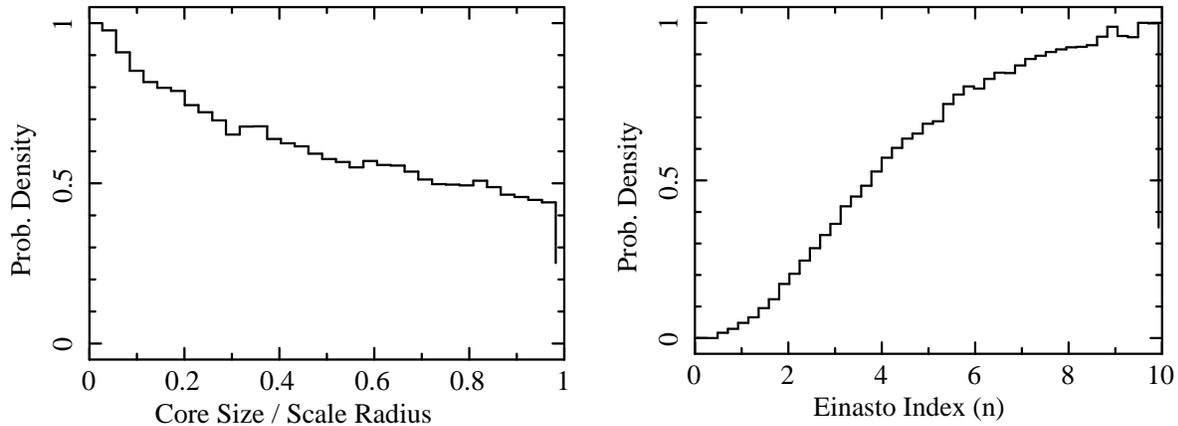

\begin{center}
\rotatebox{270}{\includegraphics[height=0.45\hsize]{coreradius.ps}}
\rotatebox{270}{\includegraphics[height=0.45\hsize]{einastoindex.ps}}
\caption {
These figures plot the posteriors for the scaled core radius of the cored NFW model ($r_c/r_S$) and the Einasto index of the Einasto model ($n$).  Interestingly, the constraints on the common halo shape parameters for the cored NFW and Einasto models indicate a more cuspy common halo shape.  However, this study does not allow for different halo profile shapes among the galaxies, but rather imposes a common shape among the full sample.  Thus, these results should be viewed as the aggregate solution to the full sample rather than a statement about any individual halo shape.  Even so, this is an interesting result given recent literature (see Section \ref{section:discussion}).
\label{fig:cuspcoreprior}
}
\end{center} 
\end{figure*}

\begin{figure*}
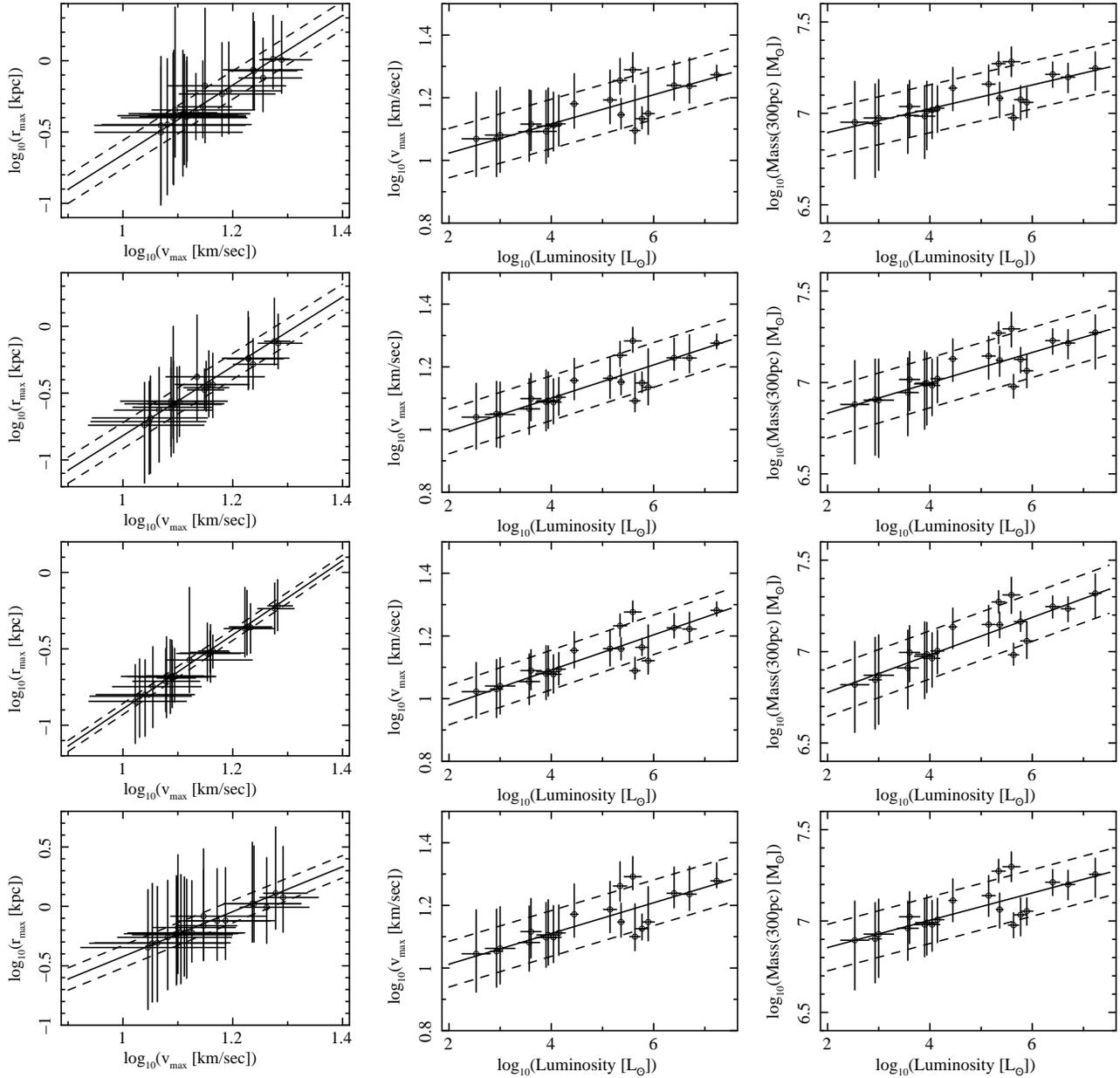

\begin{center}
\rotatebox{270}{\includegraphics[height=0.32\hsize]{gal-nfw-rv.ps}}
\rotatebox{270}{\includegraphics[height=0.32\hsize]{gal-nfw-vl.ps}}
\rotatebox{270}{\includegraphics[height=0.32\hsize]{gal-nfw-ml.ps}}
\rotatebox{270}{\includegraphics[height=0.32\hsize]{gal-corenfw-rv.ps}}
\rotatebox{270}{\includegraphics[height=0.32\hsize]{gal-corenfw-vl.ps}}
\rotatebox{270}{\includegraphics[height=0.32\hsize]{gal-corenfw-ml.ps}}
\rotatebox{270}{\includegraphics[height=0.32\hsize]{gal-burkert-rv.ps}}
\rotatebox{270}{\includegraphics[height=0.32\hsize]{gal-burkert-vl.ps}}
\rotatebox{270}{\includegraphics[height=0.32\hsize]{gal-burkert-ml.ps}}
\rotatebox{270}{\includegraphics[height=0.32\hsize]{gal-einasto-rv.ps}}
\rotatebox{270}{\includegraphics[height=0.32\hsize]{gal-einasto-vl.ps}}
\rotatebox{270}{\includegraphics[height=0.32\hsize]{gal-einasto-ml.ps}}
\caption {
This figure shows the various bottom-level posteriors for $r_{\textrm{max}}$, $v_{\textrm{max}}$, and $M(300)$.  Shown in each column, from left to right, are the individual galaxies posteriors of $\log(r_{\textrm{max}})$ versus $\log(v_{\textrm{max}})$, $\log(v_{\textrm{max}})$ versus $\log(L)$, and $\log(M(300))$ versus $\log(L)$.  From top to bottom, each row contains the posteriors assuming the NFW, cored NFW, Burkert and Einasto models.  Overlaid is the median fit prior distribution showing the distribution peak (solid line) and intrinsic one sigma variance (dashed line).  These plots show that individual posterior constraints for each galaxy agrees well with the inferred overall galaxy distribution.  The $\log(r_{\textrm{max}})$ versus $\log(v_{\textrm{max}})$ plots show the net effect of the `$\alpha_{lv}$ -- $v_{\beta_{lv}}$ degeneracy' in the extreme values of $\log(v_{\textrm{max}})$.  This effect is most prominent at low $v_{\textrm{max}}$ values where the posteriors widths increase the at more extreme $v_{\textrm{max}}$ values.  This is due to the scale radii being far from the stellar half-light radius -- an unfortunate byproduct of the approximate common scale shared by the Milky Way dSph galaxies.  The effect of this degeneracy also manifests at the low-luminosity end of the $\log(v_{\textrm{max}})$ versus $\log(L)$ relation.  But this effect is minimal compared to the overall effect on the $\log(r_{\textrm{max}})$ versus $\log(v_{\textrm{max}})$ relation.  Most notably though is the implied $\log(M(300))$ versus $\log(L)$ relation.  While this relation is fairly constant, there is a definite implied small positive slope consistent with simulated value of $0.088 \pm 0.024$ \citep{Rashkov2012}.  Note that these bottom-level posteriors contain information of both the individual galaxy fit as well as the fit of the full data set to the lower-level prior.  Thus, the width of the posteriors reflect both the uncertainty of the individual galaxy parameters as well as the quality of fit of the lower-level prior.  Models that produce distributions that fit the lower-prior well allow for a larger range in the lower-level parameters since these models naturally produce more solutions that are a good overall fit to the data.  Conversely, models that produce distributions that poorly fit the lower-level prior allow a shorter range in the lower-level posteriors for the same reason.  
Since the posteriors contain information about the full parameter space, the posterior width is the result of both the individual galaxy distribution as well as the allowed range due to the fit of the prior distribution.  Thus, a narrower posterior width is not necessarily indicate a better overall fit.
\label{fig:botlvlparams}
}
\end{center} 
\end{figure*}

In Fig. \ref{fig:botlvlparams} we plot the relevant parameter constraints for each galaxy: $\log(v_{\textrm{max}})$ versus $\log(L)$, $\log(r_{\textrm{max}})$ versus $\log(v_{\textrm{max}})$, and $\log(M(300))$ versus $\log(L)$.  Overlaid is the median fit prior distribution.  These plots show that individual posterior constraints for each galaxy agree well with the inferred overall galaxy distribution.  The $\log(r_{\textrm{max}})$ versus $\log(v_{\textrm{max}})$  plots show the net effect of the `$\alpha_{lv}$ -- $v_{\beta_{lv}}$ degeneracy' in the extreme values of $\log(v_{\textrm{max}})$.  This effect is most prominent at low $v_{\textrm{max}}$ values where the posteriors widths are the largest.
This is due to the scale radii being far from the stellar half-light radius -- an unfortunate byproduct of the approximate common scale shared by the Milky Way dSph galaxies.  The effect of this degeneracy also manifests at the low-luminosity end of the $\log(v_{\textrm{max}})$--$\log(L)$ relation.  But this effect is minimal compared to the overall effect on the $\log(r_{\textrm{max}})$--$\log(v_{\textrm{max}})$ relation.  Most notable, though, is the implied $\log(M(300))$--$\log(L)$ relation.  While this relation is fairly constant, there is a definite implied small positive slope consistent with the value of $0.088 \pm 0.024$ from simulations \citep[compare to Table \ref{tab:limits} of this paper]{Rashkov2012}.  Whereas, the implied intrinsic dispersion of $\sim -0.7$ \citep[in $\log$]{Springel2008} is consistent with the value derived here (see Table \ref{tab:limits}).  The median and 68\% credible levels for the individual galaxies parameters $\log(r_{\textrm{max}})$ and $\log(v_{\textrm{max}})$, as well as the overall distribution parameters, are summarized in Tables \ref{tab:limits} and \ref{tab:params}.  
It is important to note that these bottom-level posteriors contain information of both the individual galaxy fit as well as the fit of the full data set to the lower-level prior.  Thus, the width of the posteriors reflect both the uncertainty of the individual galaxy parameters as well as the quality of fit of the lower-level prior.  Models that produce distributions that fit the lower-level prior well allow for a larger range in the lower-level parameters since these models naturally produce more solutions that are a good overall fit to the data.  Conversely, models that produce distributions that poorly fit the lower-level prior allow a shorter range in the lower-level posteriors for the same reason.  
Since the posteriors contain information about the full parameter space, the posterior width is the result of both the individual galaxy distribution as well as the allowed range due to the fit of the prior distribution.  Thus, a narrower posterior width is not necessarily indicating a better overall fit.
This is indeed the case with the Einasto profile that even with narrower posteriors, the Bayes factor (see Table \ref{tab:evidence}) indicates that this profile is disfavoured relative to the other, more cuspy, profiles.
Also, from Fig. \ref{fig:cuspcoreprior}, we see that the constraints on the common halo shape parameters for the cored NFW and Einasto models indicated a more cuspy common halo shape as well.  However, this study does not allow for different density profile shapes among the galaxies, but rather imposes a common shape among the full sample.  Nevertheless, this result may have interesting implications on galaxy formation as discussed in the next section.

\section{Discussion}
\label{section:discussion}

Our result of a more cuspy halo, as compared to very core-like (such as the Einasto or profiles with a very shallow slope), has interesting implications given recent literature \citep{Evans2009, Strigari2010, Walker2011, Agnello2012, BreddelsHelmi2013}.  First, it is important to note that these results should be viewed as the aggregate solution to the full sample rather than a statement about any individual halo shape.  It is entirely possible that, while taken as a whole, these galaxies prefer a cuspy dark matter profile, an individual galaxy's profile may indeed exhibit more core-like behavior.  This is especially true for the more extended (and luminous) galaxies whose half-light radii do not probe the very inner portions of their dark matter distributions.  If this hypothesis is correct, it may suggest that these haloes' once cuspy inner regions were softened due to subsequent astrophysical interactions.
Another possibility is that these recent slope measurements suffer from a constant systematic bias due to asphericity in the shape of the stellar and dark matter density profiles.  A stellar profile of high asphericity can shift the measured mass at the half-light radius and implied slope by a factor of a few \citep{Kowalczyk2013}.  However, it is expected that reliable lower-limits to the inner slope can be achieved regardless of dark matter halo triaxiality \citep{Laporte2013}.  Because these haloes would be randomly originated within the sample, the bias due to triaxiality is not expected to sensitively affect central prior measurements.
Thus, the expected net effect on hierarchical modelling is an increase in the dispersion of the priors.  

This work utilizes only the line-of-sight second order velocity moments (dispersion) of a single population.  With only line-of-sight dispersions of a single population, it is not possible to uniquely determine the full mass profile \citep{Walker2009a, Wolf2010}.  So in this sense, when analysing a single galaxy, this data cannot distinguish between different mass profile models.  This work attempts to alleviate this by including information of the full galaxy sample.  But, if more information beyond the line-of-sight velocity dispersions were included, such as higher order velocity moments or multiple populations, it may be possible to individually constrain the inner slope solely by analysing a single galaxy \citep{Walker2011}.  Specifically, incorporating multiple population into this analysis would likely increase the `average' profile constraint and is subject of future work.  Other authors have obtained mass profile constraints from the use of higher order moments [for example, using methods such as phase space, Schwarzschild, or higher-order Jeans modelling \citep{Lokas2005, Wu2007, Amorisco2012, Jardel2012, Richardson2012, Breddels2013, Jardel2013}].  But these methods are more sensitive to membership issues due to foreground contamination and any physical process that would affect higher order moments \citep[e.g. unresolved binaries;][]{Minor2010}.  Thus, in ordered to avoid these systematics, we avoid such methodologies in this study.

Finally, we should mention that, while our results are consistent with a shared cuspy halo profile, there is a small preference for shallow core-like properties towards the centre.  With the exception of the Einasto profile, the Bayes factor (Table \ref{tab:evidence}) between the various models indicate that, while not definitively preferring one model, models with core-like behaviour (especially the cored NFW) have a small preference over the NFW model.  However, even if these galaxies' inner regions truly are shallow, their outer regions are very NFW-like.

The primary benefit of hierarchical modelling to dynamical mass modelling is the inclusion of relevant galaxy population distribution information.  This is particularly true in prior dominated systems in which exclusion of such information will lead to posteriors dominated by prior assumptions rather than by data.  This is the origin in the apparent difference between the results presented here and those presented in \citet{Strigari2007}.  Because some of the galaxies are prior-dominated, use of an arbitrary (non-hierarchical) prior can drastically shift the results. 
In some instances, especially with the prior-dominated ultrafaint galaxies, the differences in these posteriors differ by more than two standard deviations.  This is a typical consequence of applying a arbitrary prior instead of constraining these priors from the complete data set.  Specifically this can be seen at the high and low ends of the luminosity function.  With the use of the data-driven priors used here, we generally find lower $r_{\textrm{max}}$ and $v_{\textrm{max}}$ values at low luminosities whereas these values were a bit higher at high luminosities (when compared to \citet{Strigari2007}).

While hierarchical modelling has done quite a lot reconciling the Local Group dwarf spheriodal distribution with that predicted by CDM at the lower end of the mass function, it has only exacerbated the Too-Big-To-Fail problem.  Here we have shown that, not only would the local dwarfs have concentrations inconsistent with CDM if they were hosted by haloes with a $v_{\textrm{max}}$ of $20-40$ km/sec, but that it is {\em statistically} inconsistent for them to be hosted by haloes of a $v_{\textrm{max}}$ this size.  In other words, the improved results from this analysis undoubtedly shows that there is a deficit of Milky Way haloes with a $v_{\textrm{max}}$ of $20-40$ km/sec.

We have shown here that MLM may drastically improve results and can be applied to any problem that involves an ensemble of data sets -- given that this ensemble originate from the same underlying distribution.  Problems that meet this criterion, such as determining the mass and period distribution of planets \citep{Kepler2012} or the stellar initial mass function of star clusters \citep{Bastian2010} stand to benefit from this type of analysis.  In particular, Milky Way dSph measurements stand to benefit from this analysis because they not only meet this criterion, but previous analyses suggest that these galaxies truly follow an underlying distribution \citep{Diemand2007, Neto07, Strigari2007, Springel2008}.  Most notable, the galaxies at ultrafaint luminosities have benefited the most from this analysis.  This is entirely due to the fact that the uncertainty is simply an representation of our {\em total} `degree of belief'.  Here, it is important to realize that this belief is based on all available information:  information contained in the individual data set {\em and} information contained in the underlying distribution.  Since these galaxies' constraints on their mass profiles are dominated by the lack of knowledge of the underlying distribution, it is these constraints that have the most to gain from this methodology.  These galaxies will also be affected most if our knowledge of the underlying distribution changes.  Indications that some of the ultrafaint satellites may have dispersion lower than previously measured \citet[e.g.]{Kirby2013} would not only affect the mass measurements of those individual galaxies, but also all the ultrafaint galaxies as a whole.  However, such indications would only serve to exclude solutions with high concentrations (e.g. large $\log(r_{\textrm{max}})$--$\log(r_{\textrm{max}})$ slopes, see Fig. \ref{fig:toplvlparams}) that are inconsistent with CDM simulations.   Therefore, if this were indeed the case, we suspect that this would only strengthen our main conclusion that the Milky Way dSph's distribution is consistent within the $\Lambda$CDM paradigm for haloes with $v_{\textrm{max}} < 20$ km/sec.   

We reiterate that these conclusions are based on data constraints rather than prior assumptions.  Although it may seem that the subjectivity increased with the inclusion of seven new top-level priors, this is indeed not the case because the 20 lower-level priors are being interpreted as actual physical distributions.  As a matter of fact, the only assumption MLM requires is that the total galaxy set samples an overall distribution described by the prior probability.  But this is equivalent to assuming that each individual data set samples a larger distribution of data described by the likelihood -- an assumption that is necessary to perform any likelihood analysis.  Furthermore, the issues that plague a normal likelihood analysis also hold for the complete set of lower-level posteriors.  One example that is often overlooked is the effect of the choice of parametrization of the likelihood.  If a likelihood is parametrized with too few parameters, relevant detail may be lost or misinterpreted, whereas too many parameters may cause over-fitting.  Likewise, prior parametrization (e.g. Equation \ref{eq:botprior}) is also an issue for MLM for the same reasons.  An interesting direction of this work is to explore varying forms of $\mathcal{P}(v_{\textrm{max}}|L)$ (e.g. $\mathcal{P}(M(300)|L)$).  Our main motivation in the selection of a lognormal form of $\mathcal{P}(v_{\textrm{max}}|L)$ was the apparent flat $M(300)$-$L$ relation claimed by \citet{Strigari07-redef}.  Specifically, we questioned whether posteriors derived using this prior information would yield the same results.  Of course, we found that they did not.  But, from Fig. \ref{fig:botlvlparams} it is conceivable that the high luminosity galaxies follow a different $M(300)$--$L$ relationship than low luminosity galaxies.  If so, this may have profound consequences on our conclusions considering that the main reason for our improved constraints is the addition of luminosity information.  While it is hard to surmise what effect, if any, these issues have on the allowed density profiles and the subsequently constrained $r_{\textrm{max}}$--$v_{\textrm{max}}$ distribution, study of alternative prior forms may give insights to the consistency of alternative dark matter theories.  Because of the large dimensional parameter space that is usually involved in Bayesian hierarchical modelling we have taken advantage here of every technical simplification available.  Specifically, we have integrated over the luminosities of each individual galaxy to reduce the parameter space by a third.  But if more complicated luminosity mass relations were considered, then the integral over luminosity would not be analytic.  For this reason, these technical difficulties preclude us from including these issues in this current study.  Thus, we leave this analysis to future work.

\section{Conclusion}

In this work we introduced a new methodology to derive mass profiles for the Milky Way Local Group dwarf dSphs.  This new methodology, based on MLM, exploits the fact that these individual galaxies sample an underlying distribution.  By simultaneously constraining both the individual galaxy and the overall galaxy distribution parameters, not only do individual galaxies posteriors become more robust, but the overall distribution properties may also be inferred.  This is done by interpreting the individual galaxy prior probabilities as the probability of observing a particular galaxy from this overall distribution.  In other words, we interpret the prior probability as a {\em frequentist} probability sampling from an actual physical galaxy distribution.  Thus, in much the same way that single data point probabilities can be combined to form a `lower-level' likelihood that can be used to constrain individual galaxy parameters, the posteriors from the full galaxy sample can be combined to form a `upper-level' likelihood that can be used to constrain {\em both} the overall galaxy distribution {\em and} the individual galaxy parameters.  This interpretation of the individual prior distributions then becomes beneficial for the following reasons.  First, it removes the subjectivity normally associated with the Bayesian `degree of belief' interpretation of probability on the lower-level priors.  And secondly, it allows us to use of the combined data set to directly constrain the prior probabilities via this newly defined `higher-level' likelihood.  Because the Bayesian interpretation of probability has not been used in the formation of this likelihood, multilevel analysis can be done in both the frequentist and Bayesian frameworks.  For this analysis, we utilized the Bayesian multilevel methodology, Bayesian hierarchical modelling.  

Application of this methodology resulted not only in more robust individual galaxy mass profile constraints, but also in fairly robust constraints on the overall distribution.  The galaxies that benefited the most from this analysis were the ultrafaint dwarf satellite galaxies.  These galaxies, because of their extreme prior dominance, had the most to gain from the extra information gained from constraining the prior PDF.  Although the anisotropy-mass degeneracy was greatly minimized, we found that this indirectly caused a somewhat constrained degeneracy between the slope and intercept of the overall distribution's $r_{\textrm{max}}$--$v_{\textrm{max}}$ relation.  Even so, the overall inferred relationship between  $r_{\textrm{max}}$ and $v_{\textrm{max}}$ as well as the inferred relationship between the mass within 300$pc$ and luminosity are in excellent agreement with CDM simulations.  Also, we found that a cuspy halo is a good `average' fit to the Milky Way satellites density profiles.  Although this does not exclude the possibility that individual galaxies (especially extended high-luminosity satellites) from having cored dark matter profiles, it may suggest that these galaxies' central regions were once cusped but may have `softened' due to astrophysical interactions.
\section{acknowledgements}
GDM is very appreciative to Alex Drlica-Wagner, Jan Conrad, Louie Strigari, Annika Peter, Tuan Do, Manoj Kaplinghat, and Christian Farnier for useful comments and discussions.  GDM also acknowledges support from the Wennergren foundation.
%
\renewcommand{\thefootnote}{\fnsymbol{footnote}}
\begin{table*}
\caption{\label{tab:limits}Summary of unrestrained top-level model parameters and results}
\scriptsize
\begin{center}
\begin{tabular}{@{}cccccccl}
\noalign{\hrule height 1pt}
\vspace{1pt}
Prior & Prior & Derived value & Derived value & Derived value & Derived value \\
parameters  & range & (NFW) & (Cored NFW) & (Burkert) & (Einasto) & Description\\ 
\hline
\vspace{1pt}
$\log_{10}(\sigma_{rv})$					& [$-10$, $3$] 	& $-1.00_{-0.37}^{+0.36}$
										& $-1.01_{-0.44}^{+0.37}$ 
										& $-1.44_{-0.38}^{+0.54}$ 
										& $-1.03_{-0.61}^{+0.40}$ 
										& (log) dispersion of the $\log(r_{\text{max}})$--$\log(v_{\text{max}})$ relation\\ 
$\alpha_{rv}$		& [$-30$, $30$] 					& $2.44_{-1.20}^{+1.30}$ 
										& $2.59_{-1.10}^{+1.43}$ 
										& $2.43_{-0.95}^{+0.97}$ 
										& $1.89_{-1.23}^{+1.57}$ 
										& Slope of the $\log(r_{\text{max}})$--$\log(v_{\text{max}})$ relation\\ 
$\beta_{rv}$		& [$-30$, $30$] 					& $-3.10_{-1.58}^{+1.46}$ 
										& $-3.41_{-1.74}^{+1.27}$ 
										& $-3.32_{-1.13}^{+1.07}$ 
										& $-2.31_{-1.84}^{+1.47}$ 
										& Intercept of the $\log(r_{\text{max}})$--$\log(v_{\text{max}})$ relation\\
$\log_{10}(\sigma_{vl})$					& [$-4$, $3$] 	& $-1.10_{-0.16}^{+0.16}$ 
										& $-1.15_{-0.15}^{+0.15}$ 
										& $-1.20_{-0.13}^{+0.12}$ 
										& $-1.14_{-0.16}^{+0.16}$ 
										& (log) dispersion of the $\log(v_{\text{max}})$--$\log(L)$ relation\\ 
$\alpha_{vl}$		& [$-10$, $10$] 					& $0.05_{-0.04}^{+0.03}$ 
										& $0.05_{-0.03}^{+0.03}$ 
										& $0.06_{-0.02}^{+0.02}$ 
										& $0.05_{-0.03}^{+0.03}$ 
										& Slope of the $\log(v_{\text{max}})$--$\log(L)$ relation\\ 
$\beta_{vl}$		& [$-10$, $10$] 					& $0.93_{-0.18}^{+0.24}$ 
										& $0.89_{-0.16}^{+0.16}$ 
										& $0.87_{-0.13}^{+0.14}$ 
										& $0.91_{-0.18}^{+0.20}$ 
										& Intercept of the $\log(v_{\text{max}})$--$\log(L)$ relation\\
$\alpha_{l}$			& [$-3$, $3$] 					& $-0.07_{-0.09}^{+0.08}$ 
										& $-0.07_{-0.08}^{+0.08}$ 
										& $-0.06_{-0.08}^{+0.08}$ 
										& $0.00_{-0.06}^{+0.06}$ 
										& Slope of the luminosity function\\
$r_c/r_S$	& [$0$, $1$]							& -- & $0.40_{-0.30}^{+0.39}$ & -- & -- & Scaled core radius\\
$n$	& [$0.5$, $10$]								& -- & -- & -- & $6.87_{-2.72}^{+2.18}$ & Einasto index
\vspace{1pt}\\
\noalign{\hrule height 1pt}
\\
\noalign{\hrule height 1pt}
\vspace{1pt}
Derived prior & & Derived value & Derived value & Derived value & Derived value \\
parameters & & (NFW) & (Cored NFW) & (Burkert) & (Einasto) & Description\\ 
\hline
\vspace{1pt}
$\log_{10}(\sigma_{ml})$ &	& $-0.88_{-0.13}^{+0.14}$ 
											& $-0.86_{-0.13}^{+0.15}$ 
											& $-0.88_{-0.12}^{+0.12}$ 
											& $-0.90_{-0.11}^{+0.13}$ 
											& (log) dispersion of the $\log(M_{300})$--$\log(L)$ relation\\ 
$\alpha_{ml}$ &						& $0.07_{-0.06}^{+0.07}$ 
											& $0.08_{-0.05}^{+0.06}$ 
											& $0.10_{-0.05}^{+0.06}$ 
											& $0.07_{-0.05}^{+0.06}$ 
											& Slope of $\log(M_{300})$--$\log(L)$ relation\\ 
$\beta_{ml}$ &						& $6.76_{-0.38}^{+0.32}$ 
											& $6.67_{-0.35}^{+0.31}$ 
											& $6.57_{-0.32}^{+0.32}$ 
											& $6.71_{-0.32}^{+0.28}$ 
											& Intercept of $\log(M_{300})$--$\log(L)$ relation\\ 
\vspace{1pt}\\
\noalign{\hrule height 1pt}
\end{tabular}
\\
\end{center}
\begin{flushleft}
{Note:  All $v_{\textrm{max}}$ in km/s.  All $r_{\textrm{max}}$ and $r_{1/2}$ in kpc.  All Masses in $M_{\bigodot}$.  All Luminosities in $L_{\bigodot}$}.  All $\log$ to the base of $10$.
\end{flushleft}
\end{table*}
\begin{table*}
\caption{\label{tab:params}Summary of galaxy model parameters and results}
\scriptsize
\begin{center}
\begin{tabular}{@{}ccccccccc}
\noalign{\hrule height 1pt}
\vspace{1pt}
 & \multicolumn{2}{c}{NFW} & \multicolumn{2}{c}{Cored NFW} & \multicolumn{2}{c}{Burkert} &  \multicolumn{2}{c}{Einasto} \\
Galaxy  	& $\log_{10}(v_{\text{max}})$\footnotemark[1]
				& $\log_{10}(r_{\text{max}})$\footnotemark[2]
				& $\log_{10}(v_{\text{max}})$\footnotemark[1]
				& $\log_{10}(r_{\text{max}})$\footnotemark[2]
				& $\log_{10}(v_{\text{max}})$\footnotemark[1]
				& $\log_{10}(r_{\text{max}})$\footnotemark[2]
				& $\log_{10}(v_{\text{max}})$\footnotemark[1]
				& $\log_{10}(r_{\text{max}})$\footnotemark[2]\\
\hline
\vspace{1pt}
Carina 					& $1.09_{-0.04}^{+0.14}$ & $-0.35_{-0.33}^{+0.72}$ 
							& $1.09_{-0.03}^{+0.09}$ & $-0.58_{-0.26}^{+0.58}$
							& $1.09_{-0.03}^{+0.04}$ & $-0.69_{-0.15}^{+0.25}$ 
							& $1.10_{-0.04}^{+0.10}$ & $-0.18_{-0.39}^{+0.61}$ \\ 
Draco 					& $1.26_{-0.04}^{+0.07}$ & $-0.12_{-0.24}^{+0.28}$ 
							& $1.24_{-0.04}^{+0.04}$ & $-0.28_{-0.19}^{+0.19}$
							& $1.23_{-0.03}^{+0.04}$ & $-0.36_{-0.17}^{+0.16}$ 
							& $1.26_{-0.05}^{+0.08}$ & $-0.01_{-0.30}^{+0.42}$ \\ 
Fornax 					& $1.27_{-0.02}^{+0.03}$ & $0.01_{-0.21}^{+0.31}$ 
							& $1.28_{-0.02}^{+0.03}$ & $-0.11_{-0.18}^{+0.32}$
							& $1.28_{-0.02}^{+0.03}$ & $-0.22_{-0.16}^{+0.17}$
							& $1.28_{-0.02}^{+0.06}$ & $0.11_{-0.30}^{+0.56}$ \\
Leo I 					& $1.24_{-0.05}^{+0.09}$ & $-0.06_{-0.28}^{+0.40}$ 
							& $1.23_{-0.05}^{+0.07}$ & $-0.24_{-0.24}^{+0.35}$ 
							& $1.22_{-0.04}^{+0.05}$ & $-0.37_{-0.19}^{+0.27}$
							& $1.24_{-0.05}^{+0.09}$ & $0.02_{-0.33}^{+0.52}$ \\ 
Leo II 					& $1.15_{-0.07}^{+0.15}$ & $-0.18_{-0.40}^{+0.54}$ 
							& $1.13_{-0.06}^{+0.12}$ & $-0.38_{-0.33}^{+0.46}$
							& $1.12_{-0.04}^{+0.11}$ & $-0.57_{-0.21}^{+0.48}$ 
							& $1.15_{-0.06}^{+0.11}$ & $-0.08_{-0.37}^{+0.57}$ \\ 
Sculptor 				& $1.24_{-0.05}^{+0.08}$ & $-0.07_{-0.27}^{+0.35}$ 
							& $1.23_{-0.04}^{+0.06}$ & $-0.24_{-0.22}^{+0.31}$
							& $1.23_{-0.03}^{+0.05}$ & $-0.36_{-0.18}^{+0.24}$
							& $1.24_{-0.05}^{+0.08}$ & $0.02_{-0.32}^{+0.49}$ \\
Sextans 				& $1.13_{-0.03}^{+0.04}$ & $-0.36_{-0.19}^{+0.28}$ 
							& $1.15_{-0.03}^{+0.04}$ & $-0.48_{-0.15}^{+0.16}$
							& $1.16_{-0.03}^{+0.03}$ & $-0.51_{-0.11}^{+0.10}$
							& $1.13_{-0.04}^{+0.05}$ & $-0.22_{-0.25}^{+0.44}$ \\
Ursa Minor 			& $1.29_{-0.04}^{+0.05}$ & $0.01_{-0.24}^{+0.27}$ 
							& $1.28_{-0.04}^{+0.04}$ & $-0.13_{-0.19}^{+0.22}$
							& $1.28_{-0.03}^{+0.04}$ & $-0.24_{-0.17}^{+0.17}$
							& $1.29_{-0.04}^{+0.06}$ & $0.08_{-0.29}^{+0.43}$ \\
Bootes 					& $1.18_{-0.08}^{+0.10}$ & $-0.23_{-0.30}^{+0.36}$ 
							& $1.16_{-0.06}^{+0.07}$ & $-0.43_{-0.24}^{+0.25}$
							& $1.15_{-0.06}^{+0.06}$ & $-0.53_{-0.18}^{+0.20}$
							& $1.17_{-0.07}^{+0.10}$ & $-0.12_{-0.33}^{+0.44}$ \\
Canes Venatici I 	& $1.15_{-0.05}^{+0.05}$ & $-0.32_{-0.22}^{+0.32}$ 
							& $1.15_{-0.04}^{+0.04}$ & $-0.46_{-0.18}^{+0.20}$
							& $1.16_{-0.03}^{+0.04}$ & $-0.52_{-0.13}^{+0.14}$
							& $1.15_{-0.05}^{+0.06}$ & $-0.16_{-0.29}^{+0.47}$ \\
Canes Venatii II 	& $1.09_{-0.10}^{+0.13}$ & $-0.39_{-0.48}^{+0.46}$ 
							& $1.09_{-0.09}^{+0.10}$ & $-0.56_{-0.41}^{+0.33}$ 
							& $1.08_{-0.08}^{+0.09}$ & $-0.68_{-0.27}^{+0.26}$
							& $1.10_{-0.10}^{+0.11}$ & $-0.23_{-0.43}^{+0.49}$ \\
Coma Berentices 	& $1.09_{-0.09}^{+0.13}$ & $-0.40_{-0.46}^{+0.46}$ 
							& $1.07_{-0.08}^{+0.10}$ & $-0.63_{-0.39}^{+0.32}$
							& $1.05_{-0.07}^{+0.09}$ & $-0.75_{-0.27}^{+0.26}$
							& $1.08_{-0.09}^{+0.12}$ & $-0.26_{-0.44}^{+0.48}$ \\
Hercules 				& $1.11_{-0.08}^{+0.11}$ & $-0.37_{-0.36}^{+0.42}$ 
							& $1.09_{-0.07}^{+0.07}$ & $-0.61_{-0.29}^{+0.30}$ 
							& $1.08_{-0.06}^{+0.06}$ & $-0.71_{-0.20}^{+0.21}$
							& $1.10_{-0.08}^{+0.10}$ & $-0.24_{-0.40}^{+0.49}$ \\
Leo IV 					& $1.11_{-0.10}^{+0.12}$ & $-0.37_{-0.44}^{+0.44}$ 
							& $1.09_{-0.09}^{+0.09}$ & $-0.58_{-0.37}^{+0.32}$ 
							& $1.09_{-0.08}^{+0.08}$ & $-0.68_{-0.24}^{+0.24}$
							& $1.11_{-0.10}^{+0.11}$ & $-0.22_{-0.43}^{+0.49}$ \\
Leo T 					& $1.19_{-0.08}^{+0.10}$ & $-0.21_{-0.30}^{+0.34}$ 
							& $1.16_{-0.06}^{+0.07}$ & $-0.44_{-0.25}^{+0.23}$
							& $1.16_{-0.06}^{+0.06}$ & $-0.53_{-0.19}^{+0.19}$
							& $1.19_{-0.07}^{+0.09}$ & $-0.12_{-0.33}^{+0.45}$ \\
Segue I 				& $1.07_{-0.12}^{+0.15}$ & $-0.50_{-0.51}^{+0.46}$ 
							& $1.04_{-0.10}^{+0.11}$ & $-0.74_{-0.43}^{+0.32}$
							& $1.02_{-0.08}^{+0.09}$ & $-0.84_{-0.28}^{+0.24}$
							& $1.05_{-0.12}^{+0.14}$ & $-0.35_{-0.52}^{+0.49}$ \\
Ursa Major I 			& $1.12_{-0.07}^{+0.09}$ & $-0.38_{-0.33}^{+0.41}$ 
							& $1.10_{-0.06}^{+0.06}$ & $-0.56_{-0.26}^{+0.26}$
							& $1.09_{-0.05}^{+0.05}$ & $-0.68_{-0.17}^{+0.18}$
							& $1.11_{-0.07}^{+0.09}$ & $-0.22_{-0.36}^{+0.47}$ \\
Ursa Major II 		& $1.12_{-0.08}^{+0.11}$ & $-0.39_{-0.35}^{+0.39}$ 
							& $1.10_{-0.07}^{+0.08}$ & $-0.58_{-0.29}^{+0.28}$
							& $1.09_{-0.06}^{+0.07}$ & $-0.68_{-0.20}^{+0.20}$
							& $1.12_{-0.08}^{+0.11}$ & $-0.22_{-0.38}^{+0.45}$ \\
Willman 1 			& $1.08_{-0.12}^{+0.15}$ & $-0.45_{-0.49}^{+0.46}$ 
							& $1.05_{-0.11}^{+0.10}$ & $-0.71_{-0.39}^{+0.30}$
							& $1.04_{-0.09}^{+0.09}$ & $-0.80_{-0.27}^{+0.23}$
							& $1.06_{-0.11}^{+0.13}$ & $-0.31_{-0.49}^{+0.47}$ \\
Segue 2 				& $1.07_{-0.12}^{+0.15}$ & $-0.45_{-0.52}^{+0.48}$ 
							& $1.05_{-0.11}^{+0.10}$ & $-0.69_{-0.41}^{+0.31}$
							& $1.03_{-0.09}^{+0.09}$ & $-0.81_{-0.27}^{+0.24}$
							& $1.05_{-0.11}^{+0.13}$ & $-0.31_{-0.50}^{+0.50}$ 
\vspace{1pt}\\
\noalign{\hrule height 1pt}
\end{tabular}
\end{center}
\begin{flushleft}
\footnotemark[1]{All $v_{\textrm{max}}$ in km/s.} \\
\footnotemark[2]{All $r_{\textrm{max}}$ in kpc.} 
\end{flushleft}
\end{table*}
%
\bibliographystyle{apj}

\bibliography{bib}

\end{document}